\documentclass[10pt]{article}

\usepackage{amsmath}
\usepackage{amssymb}

\usepackage{graphicx,wrapfig}

\usepackage{cite}

\usepackage{color} 

\usepackage[labelfont=bf,labelsep=period,justification=raggedright]{caption}

\makeatletter
\renewcommand{\@biblabel}[1]{\quad#1.}
\makeatother

\date{}

\usepackage{comment}
\usepackage{graphics}
\usepackage{longtable}
\usepackage{amsthm, amsfonts}
\usepackage{placeins}
\usepackage{setspace}
\usepackage{float}

\def\k#1{k_{#1}}

\def\PIPtr{$\tn{PIP}_3$}
\def\PIPtw{$\tn{PIP}_2$}

\def\etot{Ras_{total}}
\def\ea{Ras}
\def\eb{Ras^\#}
\def\ec{Ras^*}
\def\ed{Ras^{\#*}}
\def\gf{RasGEF^*}
\def\gp{RasGAP^*}
\def\Gbg{G_{\beta\gamma}}

\def\lf{\left}
\def\rt{\right}

\DeclareFontFamily{OT1}{pzc}{}
\DeclareFontShape{OT1}{pzc}{m}{it}%
              {<-> s * [1.10] pzcmi7t}{}
\DeclareMathAlphabet{\mathpzc}{OT1}{pzc}%
                                 {m}{it}

\def\c#1{\mathcal{#1}}

\def\tn#1{\textnormal{#1}}

\def\figlabel#1{{\it (#1)}}

\def\grad{\triangledown}

\def\del{\partial}

\def\ie{{\it i.e. }}

\def\tildel{{\raise.17ex\hbox{$\scriptstyle\sim$}}}

\def\delt#1{\frac{\partial #1}{\partial t}}
\def\D#1{\dfrac{d{#1}}{dt}}
\def\Dn#1{\dfrac{\del{#1}}{\del n}}

\newcommand{\stk}[2]{~\mathop{\stackrel{\rightharpoonup}{\leftharpoondown}}^{#1}_{#2}~}
\def\stki#1{~\stackrel{#1}{\longrightarrow}~}
\newcommand{\dx}[2]{\dfrac{d{#1}}{d{#2}}}
\newcommand{\Dx}[2]{\dfrac{\del{#1}}{\del{#2}}}
\newcommand{\Dxx}[2]{\dfrac{\del^2{#1}}{\del{#2}^2}}

\def\dd{{\em Dictyostelium}}

\def\genen#1{{\it #1}}

\def\citet#1{\cite{#1}}
\def\altcite#1{\cite{#1}}

\begin{document}

\begin{flushleft}
{\Large
\textbf{Signal transduction and directional sensing in
eukaryotes }
}
\\
Varunyu Khamviwath$^{1}$, 
Hans G. Othmer$^{1,2,\ast}$
\\
\bf{1} School of Mathematics, University of Minnesota, Minneapolis, Minnesota,  U.S.A.
\\
\bf{2} Digital Technology Center, University of Minnesota, Minneapolis, Minnesota,  U.S.A.
\\
$\ast$ E-mail: othmer@math.umn.edu
\end{flushleft}


\section*{Abstract} 

Control of the cytoskeleton and mechanical contacts with the extracellular
environment are essential component of motility in eukaryotic cells. In the
absence of signals, cells continuously rebuild the cytoskeleton and periodically
extend pseudopods or other protrusions at random membrane locations.
Extracellular signals bias the direction of movement by biasing the extension of
protrusions, but this involves another layer of biochemical networks for signal
detection, transduction, and control of the rebuilding of the cytoskeleton.
Here we develop a model for the latter processes that centers on a Ras-based
module that adapts to constant extracellular signals and controls the downstream
PI3K--PIP$_3$-based module responsible for amplifying a spatial gradient of the
signal. The resulting spatial gradient can lead to polarization, which enables
cells to move in the preferred direction (up gradient for attractants and
down-gradient for repellents).  We show that the model can replicate many of the
observed characteristics of the responses to cAMP stimulation for \dd, and
analyze how cell geometry and signaling interact to produce the observed
localization of some of the key components of the amplification module. We show
how polarization can emerge without directional cues, and how it interacts with
directional signals and leads to directional persistence. Since other cells such
as neutrophils use similar pathways, the model is a generic one for a large
class of eukaryotic cells.

\section*{Author Summary}

Eukaryotic cells move in response to extracellular signals in a variety of
contexts, including the immune response, the formation of vascular networks
during development, and metastasis of tumor cells in cancer. The transduction of
extracellular signals into changes in the cellular cytoskeleton, which is an
essential component of directed movement, is a complex process that involves
several layers of control that we partition into modules based on the
biochemical steps and their purpose. To enable cells to respond to a wide range
of signals cells detect changes in the signal and ignore constant background
signals, and this is encapsulated in a Ras-based module in our model. However,
extracellular signals are frequently weak, and therefore reliable control of the
motile machinery requires amplification of the extracellular signal, and this is
performed by an application module based on PI3K, a protein kinase that controls
the phosphorylation of certain membrane lipids. The model can replicate much of
the observed response of the cellular slime mold \dd\ to changes in cAMP, which
is the signaling molecule.

\section*{Introduction}

Cell and tissue movement is an integral part of many biological
processes, such as large-scale tissue rearrangements or translocations
that occur during embryogenesis, wound healing, angiogenesis, and axon
growth and migration. Individual cells such as bacteria migrate toward
better environments by a combination of taxis and kinesis, and
macrophages and neutrophils use these same processes to find bacteria
and cellular debris as part of the immune response. Our understanding
of signal transduction and motor control in flagellated bacteria such
as {\em E. coli} that move by swimming and bias their movement by
control of their run lengths is quite advanced
\cite{Othmer:2013:EAB} compared with our
understanding of how amoeboid cells such as macrophages crawl through
tissues.  The fundamental issues in the latter context include how
directional information is extracted from the extracellular signals,
how cells develop and maintain polarity, how cells exert traction on
their environment, and how adhesion to substrates or
other cells is controlled.

The cellular slime mold {\em Dictyostelium discoideum} (Dd) is an amoeboid
cell that is widely-used as a model system for studying signal
transduction, chemotaxis, and cell motility. After starvation triggers
the transition from the vegetative to the aggregation phase, Dd uses
3'-5'cyclic adenosine monophosphate (cAMP) as a messenger for
signaling by pacemaker cells to control cell movement in various
stages of development \cite{Othmer:1998:OCS}. The production and relay
of cAMP pulses by cells that are excitable but not oscillatory,
coupled with chemotactic movement toward the source of cAMP,
facilitates the organization of large territories.  In early
aggregation the cells move autonomously, but in late aggregation they
form connected streams that migrate toward the pacemaker (reviewed in 
\altcite{Othmer:1998:OCS}).  

Cell motion in Dd consists of the alternating extension of pseudopods and
retraction of trailing parts of the cell \cite{Soll:1993:MBA}.  Not all
extensions are persistent, in that they must anchor to the substrate or to
another cell, at least temporarily, in order for the remainder of the cell to
follow \cite{Soll:1995:UCU}.  In the absence of cAMP stimuli, un-polarized Dd
cells extend pseudopods in random directions, presumably in order to determine a
favorable direction in which to move. Polarized cells have a high propensity to
extend new pseudopods on alternate sides at the leading edge, which facilitates
maintenance of their direction of movement \cite{Haastert_2004, Li_2008,
Bosgraaf_2009}.  Aggregation-competent cells respond to cAMP stimuli with
characteristic changes in their morphology. The first response is suppression of
existing pseudopods and rounding up of the cell (the `cringe response'), which
occurs within about 20 s and lasts about 30 s
\cite{McRobbie:1983:CAA,Condeelis:1990:MAC}.  Under uniform elevation of the
ambient cAMP this is followed by extension of pseudopods in various directions,
and an increase in the motility \cite{Varnum:1985:DAA,Wessels:1992:BDA} and
polarity \cite{Zigmond_1977, Kriebel_2003}. A localized application of cAMP
elicits the cringe response followed by a localized extension of a pseudopod
near the point of application of the stimulus \cite{Swanson:1982:LSC}. This type
of stimulus is similar, although it varies more rapidly, to that a cell
experiences in cAMP waves during aggregation. Cells undergo periodic shape
changes from rounded to elongated in response to waves during aggregation
\cite{Alcantara:1974:SPD}, and waves elicit the cringe response
\cite{Meier_2011}.  Both polarized and un-polarized cells are able to detect and
respond to shallow chemoattractant gradients of the order of a 2\% concentration
difference between the anterior and posterior of the cell
\cite{Parent:1999:CSD}. While unpolarized cells are sensitive to directional
cues at all points along the perimeter, polarized cells are more sensitive at
their leading edge.  Directional changes of a shallow gradient induces
reorientation of polarized cells, whereas large changes in the attractant lead
to retraction of a pseudopod and formation of a new one in the direction of the
stimulus \cite{Gerisch:1982:CD, Andrew_2007}.

Cells also respond to static gradients of cAMP.  Fisher et al.~\cite{Fisher:1989:QAC} 
showed that cells move faster up a cAMP gradient than
down, and that the majority of turns made by a cell are spontaneous (although
there is a reduction in the frequency of turns when the cell moves up
the gradient). However, the magnitude and direction of a turn is strongly
influenced by the gradient in that there is a strong tendency to move up the 
gradient. This was also demonstrated under treatment of latrunculin A (latA),
where immobilized cells polarize their filamentous actin (F-actin) localization
towards a directional cue \cite{Parent:1999:CSD}.  Furthermore, aggregation is
not affected by the absence of relay (treating cells with caffeine suppresses
relay but does not impair their chemotactic ability
\cite{Brenner:1984:CBA,Siegert:1989:DIP}).

In addition to responding to changes in cAMP, due for example to
movement in a static gradient, local application of a stimulus, or the stimulus
that results from cAMP waves in aggregation fields, cells also adapt to constant
background levels of cAMP, which means that they respond to transient changes in
the stimulus, but not to constant stimuli. This ability to adapt to the mean
stimulation level over several orders of magnitude allows cells to respond to
repeated stimulation's and develop sensitivity to small difference in the cAMP
level across the cells \cite{Parent_1998, Takeda_2012}. Detailed mathematical
models based on the cAMP signal transduction pathway can reproduce this behavior
\cite{Tang_1995,Dallon_1998},  and a cartoon model that illustrates the
essential dynamics of excitation and adaptation is given in
\cite{Othmer:1998:OCS}. In any case, is should be noted that not all state
variables return to pre-stimulus levels in systems that adapt -- some state
variables must change in order to compensate for changes in the background
stimulus level \cite{Othmer:1998:OCS}.

The spatio-temporal chemotactic activities in Dd have been visualized by
localization of tagged F-actin and other molecules within the chemotactic pathway, such
as phosphatidylinositol-3,4,5-trisphosphate (\PIPtr) and active Ras. The first
phase of the response, which corresponds to 'cringing', is characterized by
uniform and transient localization of these molecules along the cell periphery
within 10 $s$. Then activity at the cell membrane drops after $30-50~s$ and is
followed by the second phase of the response that involves localized membrane
activity towards directional cues or in newly created pseudopods
\cite{Chen_2003, Sasaki_2004, Xu_2005}.

Many components in the signal transduction pathway governing chemotaxis in Dd
have been identified.  Ras is a family of small G-proteins whose two members,
RasC and RasG, are the common regulators of parallel pathways that control
chemotactic activities and are necessary for chemotaxis and relay of the cAMP
signal in Dd \cite{Kortholt_2011}. They are also the most upstream molecules
within the chemotactic pathway whose activity adapts
\cite{Zhang_2008,Kortholt_2011}. RasG is a primary regulator of
phosphatidylinositol-3 kinase (PI3K), which converts
phosphatidylinositol-4,5-diphosphate (\PIPtw) into \PIPtr, while RasC regulates
activity of the target of rapamycin complex 2 (TORC2), a parallel pathway that
regulates chemotaxis and signal relay. \PIPtr\ is a membrane lipid which
contains a specific site that binds and activates many effectors containing a
pleckstrin homology domain (PH domain). Its direct and indirect effectors
include proteins such as RacB, RacC, and WASP that lead to F-actin
polymerization, proteins such as PKB/Akt and PhdA that regulate cell polarity
and chemotaxis, and the cytosolic regulator of adenylyl cyclase (CRAC), which is
necessary for cAMP production. The PI3K activity is crucial for polarity,
chemotaxis in a shallow gradient, and stimulus-dependent increase in motility
and pseudopod generation \cite{Funamoto_2001, Andrew_2007, Afonso_2011}. Sasaki 
et al.~\cite{Sasaki_2004, Sasaki_2007} showed that there is positive feedback
between PI3K, F-actin, and Ras, and this feedback is necessary for spontaneous
generation of pseudopods in the absence of external stimuli. In fibroblasts,
PI3K stabilizes membrane protrusions. Exogenous Rac activation drives creation
of new pseudopods, but the activity is not sustained in the absence of PI3K
activity \cite{Welf_2012}.

Although the biochemistry underlying the Ras--PI3K--F-actin network has been
well-studied, the mechanism leading to spontaneous pseudopod formation and
robust biphasic responses to stimuli remains elusive. Early models addressed
cAMP relay \cite{Goldbeter_1977, Tang_1995} and spontaneous pseudopod formation
\cite{Meinhardt_1999}. More recently, a biochemical-based model
\cite{Meier-Schellersheim_2006} and an abstract model \cite{Xiong_2010,Shi_2013} that
exhibit the biphasic response have been proposed. The former was drawn from
parametric optimization on an extensive signaling network where it is difficult
to develop intuition and understand the underlying mechanisms. The latter
provided an abstract model in which the signal adapts within an upstream network
due to a feedforward mechanism. In this model detection of spatial gradients
involves an activator-inhibitor mechanism \cite{Gierer:1972:TBP}, renamed as the
local-excitation-global-inhibition (LEGI) network \cite{Xiong_2010,Shi_2013},
which is structurally similar to an earlier cAMP relay model
\cite{Tang_1995}. However the LEGI model is formal, and no attempt has been made
to identify the components with the known components of the signal transduction
network.  A recent study of Ras activity showed that adaptation occurs at this
regulation step and argued that adaptation is due to feedforward control
\cite{Takeda_2012}. In addition, there are models that address polarization and
spontaneous local \PIPtr\ activity, which may also be induced by external
stimulation \cite{Gamba_2005, Mori_2008, Hecht_2010}. Jilkine and
Edelstein-Keshet \cite{Jilkine_2011} give a more detailed account of existing
directional sensing and polarization models, but despite extensive theoretical
studies on the pathway, the roles of various molecular players in driving the
processes that lead to robust and highly sensitive directional responses and
cell polarity are not well understood. Further remarks on existing models are
relegated to the discussion.

Here we propose a model based on known biochemistry that exhibits robust
adaptation and amplification of extracellular signals, and for which the
predictions match experimental observations. It consists of two modules: (i) an
upstream module structurally similar to the biochemical model in
\cite{Takeda_2012} which regulates Ras activity by a feedforward control and is
responsible for adaptation to the mean stimulus level and (ii) a downstream
module which controls \PIPtr\ activity and amplifies subtle spatial gradients of
Ras activity. In this model, detection of spatial gradients via the upstream
module depends on the assumption that the characteristic decay length $L_d
\equiv \sqrt{D_{(\cdot)}/k_{(\cdot)}}$, wherein $(\cdot)$ denotes the species
label, of RasGAP exceeds that of RasGEF.  We later assume equal diffusion
coefficients for RasGEF and RasGAP and the numerical values used for the decay
constants lead to a ratio of $\sqrt{2}$ for the characteristic lengths. Thus
this assumption is less restrictive than the assumption on diffusion
coefficients needed in the LEGI framework, which requires the activator to be
local and the inhibitor to diffuse rapidly \cite{Levchenko_2002}.  Using this
model, we study drivers for the first and second peaks of the actin response to
stimulation, and the factors that control cell responsiveness to changes of the 
gradients and to the reapplication of uniform stimulation.   We show that cell
polarity regulates directional responses via cell geometry, and that the
internal signal integrates with extracellular stimulation via the activity of
RasGEF and RasGAP, the regulators of Ras. This unified framework exhibits
dependence of polarization degree on the level of the stimulus, infers modes of
migration, and explains directional persistence in polarized cells. Furthermore,
it suggests positive feedback between cell shape and biochemical signaling that
may lead to random pseudopodia extension and polarization, which is dependent on
PI3K activity, and a potential role of cAMP secretion in progressive development
of cell polarity through the developmental cycle.

 \subsection*{Biochemical pathways leading to $\mathbf{PIP_3}$ responses}

\begin{figure}[h] 
\centering 
\includegraphics[width=4.5in]{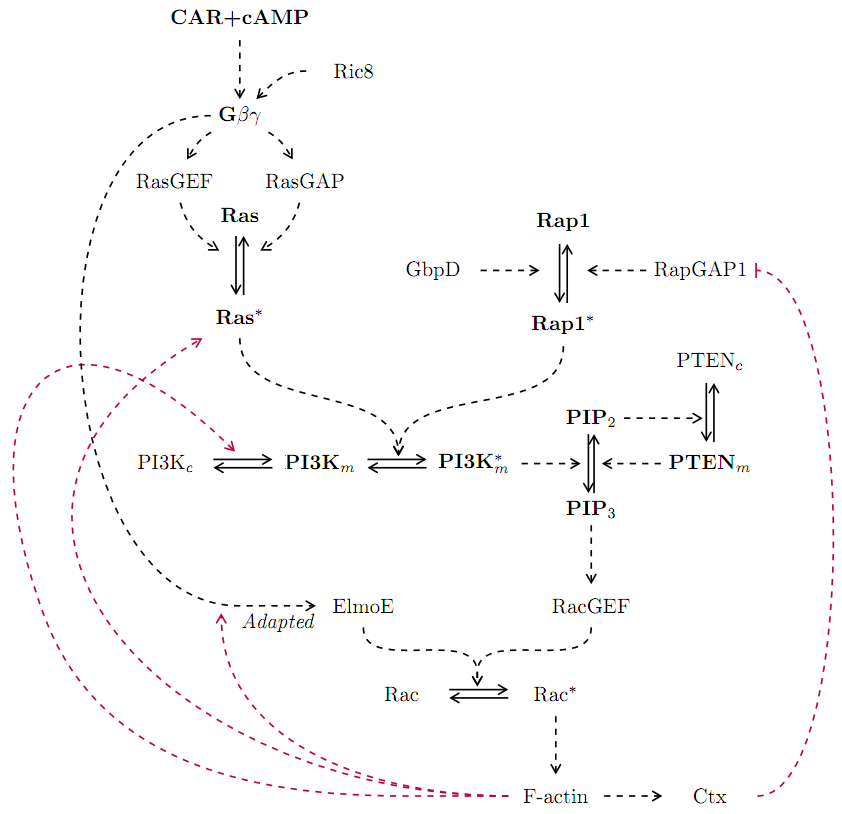} 
\caption{{\bf PI3K signaling
pathway.}  Interconversion between different forms of the same molecule is
denoted by solid arrows while positive regulation and promotion of a particular
species and process are denoted by dashed arrows.  Membrane-bound species are
shown in bold while other species reside in the cytosol. 
 \label{network}}
\end{figure}

 The first step in \dd\ chemotaxis involves binding of cAMP to CARs, G-protein coupled
receptors (GPCRs) that transduce signals by activating
heterotrimeric G proteins \cite{Kortholt_2008}. A cAMP-bound GPCR acts as a
guanine nucleotide exchange factor (GEF) for the G$_\alpha$ subunit of the
heterotrimeric G protein, causing dissociation of the activated G$_\alpha$
subunit and the G$_{\beta \gamma}$ subunit. Hydrolysis of GTP in the G$_\alpha$
subunit induces re association, which diminishes active G-protein subunits when
external cAMP is removed. Activation of the heterotrimeric G-proteins is
enhanced by resistance to inhibitors of cholinesterase 8 (Ric8), a non-receptor
independent GEF for G$_\alpha$, which is specific to free G$_\alpha$ subunits
\cite{Kataria_2013}. Free G$_{\beta \gamma}$, of which there is only one type in
\dd, is involved in activating the downstream PI3K chemotactic responses via
RasGEF and ElmoE, as shown in Figure \ref{network}.

The first module in the network involves Ras, which acts as a molecular switch
that cycles between an active GTP-bound form and an inactive GDP-bound form.
Conversion between the GTP and GDP-bound states is controlled by GTP exchange
factors (GEFs), and GTPase activating proteins (GAPs). GEF proteins activate Ras
by catalyzing the exchange of bound GDP with GTP, whereas GAPs inactivate Ras by
increasing their rate of GTP hydrolysis.  There are many GEFs in \dd, of which
RasGEFA and RasGEFR, are involved in chemotaxis. GEFs are normally in the
cytosol, but are recruited to the membrane in response to stimuli
\cite{Wilkins:2005:DGE}.  Active RasGEFA is responsible for activating RasC,
while active RasGEFR is responsible for the majority of RasG activity
\cite{Kae_2007}. RasC and RasG together are necessary for chemotactic responses
at the leading edge and the trailing edge, as well as for cAMP secretion and
cGMP production. RasC is necessary for the activity of the target of rapamycin
complex 2 (TORC2) pathway, while RasG is uniformly distributed along the plasma
membrane and directly activates PI3K \cite{Funamoto_2002, Bolourani_2008,
Charest_2010, Kortholt_2011}.  The only known RasGAP related to chemotaxis is
DdNF1, which is partly responsible for RasG deactivation, and may be involved in
detection of directed stimuli \cite{Zhang_2008}.

The activation of PI3K by active RasG depends on its localization at the
membrane, and membrane localization of PI3K depends  on F-actin activity
\cite{Funamoto_2002,Sasaki_2004}. Active PI3K phosphorylates the membrane lipid
\PIPtw\ into \PIPtr\ and the PI3K activity is balanced by phosphatase and tensin
homolog (PTEN), which is recruited to the plasma membrane by \PIPtw\ and
converts \PIPtr\ into \PIPtw.  Due to its specific PH domain, \PIPtr\ has many
downstream effectors, including RacB and possibly RacC, and the activity of both
Rac proteins ultimately leads to F-actin polymerization \cite{Park_2004,
Han_2006}.  In addition, RacB may be activated independently of the PI3K pathway
via ElmoE, which is another effector downstream of G$_{\beta \gamma}$
\cite{Yan_2012} (Figure \ref{network}).  Although F-actin is still polarized by a
cAMP gradient when PI3K activity is inhibited by LY294002, the PI3K activity is
necessary for formation of random pseudopods and for the second peak of the
F-actin activity under uniform cAMP stimulation. In fibroblasts, pseudopods
formed by photo-activation of Rac are not stabilized in the absence of PI3K
activity \cite{Funamoto_2001, Chen_2003, Loovers_2006, Sasaki_2007,Welf_2012}.

Many positive feedback steps have been identified in the PI3K pathway, including
the actin-dependent localization of PI3K at the membrane.  Moreover,
unstimulated cells, as well as \genen{g$_\beta$} null mutants, exhibit
spontaneous localization of F-actin, \PIPtr, PI3K, and Ras activity to regions
of the cell membrane that coincide with pseudopod formation. Inhibition of
either PI3K activity or F-actin leads to disruption of this spontaneous
activity, although small pseudopodial projections are observed in the absence of
PI3K activity. Interestingly, the local activity does not depend on the TORC2
pathway and substrate attachment. Moreover, disruption of RasG only leads to
mild defects in the spontaneous activity \cite{Sasaki_2004, Sasaki_2007}.  In
addition, F-actin positively regulates Rap1, a Ras-subfamily protein that
directly activates PI3K, via deactivation of RapGAP1. Ctx, an actin bundling
protein, sequesters RapGAP1 and promotes Rap1 activity at the leading edge
\cite{Kortholt_2006,Jeon_2007,Kortholt_2010}.  Studies have shown that the
distribution of cAMP receptors remains uniform under stimulation, and
localization of free G$_{\beta\gamma}$ closely follows cAMP stimulation,
suggesting that adaptation does not occur at this level \cite{Janetopoulos_2001,
Xu_2005, Xu_2007}. Since Ras activity adapts to uniform stimuli while RasGEF
remains active, adaptation of the chemotactic response probably occurs at this
step. Membrane localization of ElmoE also adapts, and since its activity is
independent of Ras activity, adaptation probably occurs here as well. From these
observations it follows that the PI3K chemotactic pathway consists of a
self-sustainable network of interconnected feedback loops whose inputs are
upstream signals that adapt at the level of Ras and ElmoE.

\begin{figure}[h] 
\centering 
\resizebox{3.5in}{!}{\includegraphics{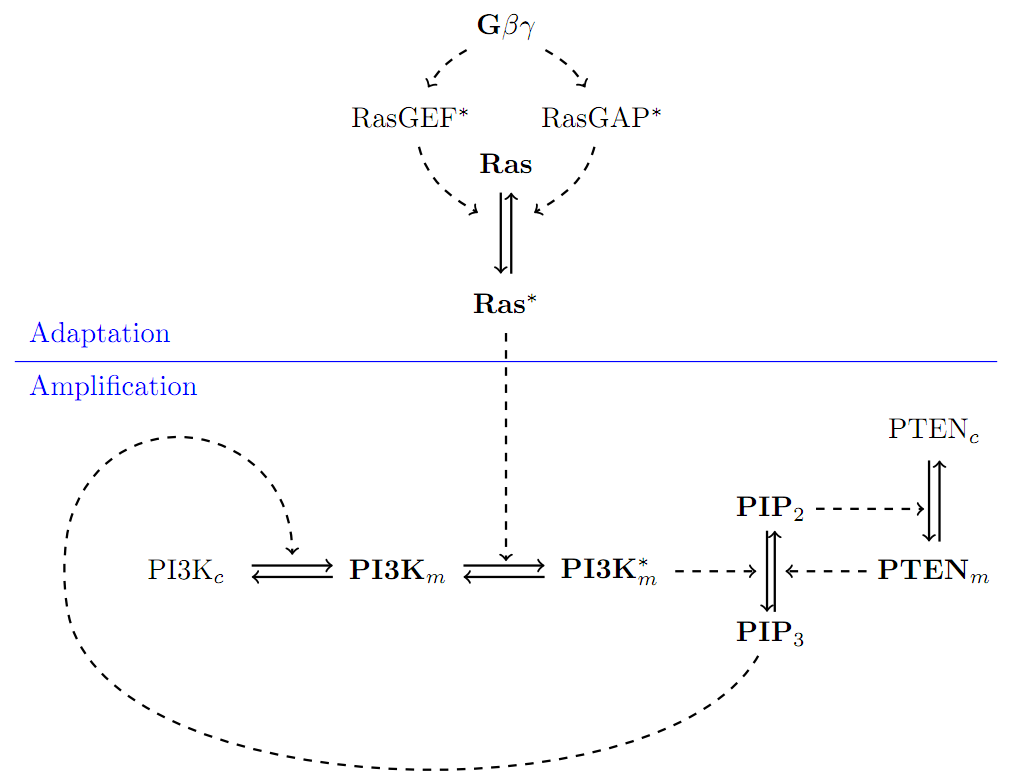}} 
\caption{{\bf A simple model of the PI3K-signaling pathway.} Membrane species are shown in
bold. 
\label{cartoonPI3K}}  
\end{figure}

To understand how the fundamental processes in the Ras and PI3K modules
contribute to cellular response to stimuli, we first study a network that 
maintains key characteristics of the full pathway, namely,  adapted input and
positive feedback. We construct a model by selecting a minimal set of 
well-understood components of the network that is capable of producing the
biphasic response that adapts to the mean stimulation level and amplifies
spatial gradients of directional cues. The model includes the activity of RasGEF,
RasGAP, Ras, PI3K, PTEN, and \PIPtr, as shown in Figure \ref{cartoonPI3K}.
Because the heterotrimeric G-protein activity closely reflects extracellular
cAMP concentration at the membrane, we identify local membrane density of free
$\Gbg$ as the input. Two-dimensional domains which represent cross sections of
cells parallel to the substrate are employed for numerical simulations. As
chemotactic responses in Dd are normally studied in latA-treated immobile cells
which assume a circular cell shape, we first study the system on a 2D disk of
$8~\mu m$ radius. Later we study the response to stimuli in more realistic cell
shapes. A detailed description of the reactions involved and the evolution
equations for the various species in  the model is given  in Methods section.

\section*{Results} 
\subsection*{Adaptation to uniform stimuli}

\begin{figure}[h] 
\centering 
\includegraphics[width=5in]{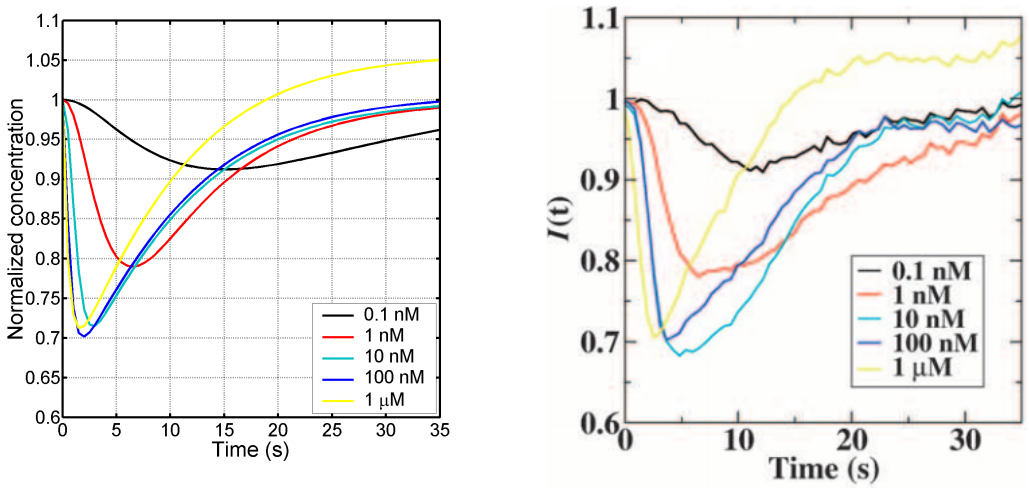} 
\caption[Ras-activation dynamics to uniform stimulation at different cAMP
levels.]{{\bf Ras-activation dynamics.} Uniform stimulation causes a transient
decrease in the average cytosolic concentration of RBD. The stimulus is applied
at $t=0~s$ and response is measured over the physiological range of cAMP
concentration. Simulation results {\it (left)} are compared to experimental
measurements {\it (right)} from \cite{Takeda_2012}. Here and hereafter simulation results 
are based on parameters in Table \ref{params}.}
\label{RasAdapt}
\end{figure}

As previously noted, Takeda et al. \cite{Takeda_2012} suggested that adaptation
in Ras activity is due to feedforward adaptation via activation and inactivation
of Ras by RasGEF and RasGAP, both of which are activated by cAMP binding to
CAR\@.  They monitored Ras activation via membrane localization of Ras-binding
domain (RBD), which diffuses freely in the cytosol and is localized to the
membrane by binding to active Ras.  We show the comparison between their
observations and our model predictions in Figure \ref{RasAdapt}, which shows the
simulated bound RBD compared with the observations. 
The authors noted that the
steady-state Ras activity monitored drops below the basal level when the system
becomes saturated at around 1 $\mu M$ cAMP, and the figure shows that the model
captures all aspects of the observed transient behavior, including more rapid
adaptation at higher stimulus levels.  Note that an undershoot of bound RBD -- 
\ie, a level below 1 -- corresponds to an overshoot of the average cytosolic
RBD shown in the figure.  Our model, which is based on the measurements of
cytosolic RBD intensity to determine rate constants given in Table \ref{params}
and used in the Ras module (Equations (\ref{Ras1})--(\ref{Ras7})), differs from those
proposed earlier \cite{Takeda_2012, Shi_2013} in that we allow both RasGEF and
RasGAP to diffuse in the cytosol and maintain conservation of these proteins. It
is precisely the conservation condition that leads to saturation of RasGEF and
RasGAP activity at a high cAMP level.
\small
\addtolength{\tabcolsep}{-2pt}
\begin{longtable}{llll}
\label{params}
 Parameter &  Value & Description & References \\	\hline
\endfirsthead
Parameter &  Value & Description & References \\	\hline
\endhead
\endfoot
\hline\\
\caption{{\bf Parameter values used in the model of the PI3K-signaling pathway.} \label{PIPparamTable}}
\endlastfoot
$R$ &   8 $\mu m$   & Cell radius     &      \\
$\delta$ & 10 $nm$     & Effective length for membrane reactions     &      \\
\hline
$RasGEF_0$ &  0.1 $\mu M$    & Average RasGEF cytosolic concentration     & \cite{Ma_2004}     \\
$RasGAP_0$ &  0.1 $\mu M$    & Average RasGAP cytosolic concentration     &  \cite{Ma_2004}    \\
$Ras_0$ &  2000 $\#/\mu m^2$     & Membrane density of Ras     &  \cite{Meier-Schellersheim_2006}    \\
$PI3K_0$ & 0.1 $\mu M$     & Average PI3K cytosolic concentration     & \cite{Ma_2004}     \\
$PTEN_0$ & 0.1 $\mu M$     & Average PTEN cytosolic concentration     & \cite{Ma_2004}     \\
$PIP_0$ & 1000 $\#/\mu m^2$     & Membrane density of PIP$_2$ and PIP$_3$     & \cite{Ma_2004,Gamba_2005}     \\
\hline
$D_{RasGEF}$ & 10 $\mu m^2/s$     & Diffusion constant for RasGEF     &  \cite{Postma_2001}    \\
$D_{RasGAP}$ &  10 $\mu m^2/s$    & Diffusion constant for RasGAP     & \cite{Postma_2001}     \\
$D_{PI3K}$ &  10 $\mu m^2/s$    & Diffusion constant for PI3K    & \cite{Postma_2001}     \\
$D_{PTEN}$ &  10 $\mu m^2/s$    &  Diffusion constant for PTEN    &  \cite{Postma_2001}    \\
\hline
$k_{RasGEF^*}$ & 93.75 $(\#/\mu m^2)^{-1}\mu m/s$    & RasGEF activation by G$\beta\gamma$    &      \\
$\k{RasGEF}$ & 0.25 $s^{-1}$     & Spontaneous RasGEF$^*$ deactivation     &      \\
$k_{RasGAP^*}$ & 1.5  $(\#/\mu m^2)^{-1}\mu m/s$   & RasGAP activation by G$\beta\gamma$     &      \\
$\k{RasGAP}$ & 0.12 $s^{-1}$     &  Spontaneous RasGAP$^*$ deactivation    &      \\
$k_{Ras^*}$ & 800 $\mu M^{-1} s^{-1}$     &  Ras activation by RasGEF$^*$    &      \\
$\k{Ras}$ & $2.5\times10^6~\mu M^{-1} s^{-1}$     & Ras$^*$ deactivation by RasGAP$^*$     &      \\
$\k{s,Ras^*}$ & $3.14\times10^{-4}~ s^{-1}$     & Spontaneous Ras activation     &      \\
$\k{s,Ras}$ & $0.03 s^{-1}$     & Spontaneous Ras$^*$ deactivation     &      \\
\hline
$k_{PI3k_m^*}$ &  18.75 $(\#/\mu m^2)^{-1}s^{-1}$    & PI3K activation by Ras$^*$     &      \\
$k_{d,PI3k_m}$ & 0.844 $s^{-1}$     & Spontaneous PI3K$^*$ deactivation     &      \\
$k_{PI3k_c}$ & $3\times10^5~s^{-1}$     & Spontaneous PI3K membrane dissociation      &      \\
$k_{PI3k_m}$ & 1500 $\mu M^{-1} s^{-1}$     & PI3K membrane binding induced by PIP$_3$    &      \\
$k_{PIP_3}$ & 720 $(\#/\mu m^2)^{-1}s^{-1}$     & PIP$_3$ production by PI3K     &      \\
$\k{PIP_2}$ & 1050 $(\#/\mu m^2)^{-1}s^{-1}$     & PIP$_3$ dephosphorylation by PTEN     &      \\
$\k{b,PI3k_m}$ &  1500 $s^{-1}$    & Spontaneous PI3K membrane binding      &      \\
$\k{PTEN_m}$ &  0.75 $\mu M^{-1} s^{-1}$   & PI3K membrane binding induced by PIP$_2$     &      \\
$k_{PTEN_c}$ &  0.375 $s^{-1}$    & Spontaneous PTEN membrane dissociation     &      \\
\end{longtable}
\addtolength{\tabcolsep}{2pt}
\normalsize

In more detail, the transient decrease in cytosolic RBD observed experimentally
corresponds to an increase in Ras activity, which is due to faster activation of
RasGEF than RasGAP.  Further, notice that the peak of Ras activity increases,
whereas the activation time and the adaptation time decrease with increases in
the cAMP level up to $\tildel10~nM$.  The model predicts a similar trend in the
response times, and shows that the rates of Ras activation and inactivation by
RasGEF and RasGAP are strongly stimulus dependent. 
Moreover, return to the basal Ras activity in the model is ensured when both
RasGEF and RasGAP activity is unsaturated, for then their steady-state
activities are proportional, which renders the steady-state Ras activity
independent of stimulation, provided spontaneous Ras activation and deactivation
are negligible.  Saturation of RasGEF and RasGAP leads to under- and
over-activitation of Ras at the steady state. In the simulation, RasGEF becomes
saturated before RasGAP, resulting in a lower Ras activity and more free
RBD.

\subsection*{Detection of spatial gradients by PI3K-$\mathbf{PIP_3}$ feedback loop}

\begin{figure}[h]
 \centering
 \includegraphics[width=5.in]{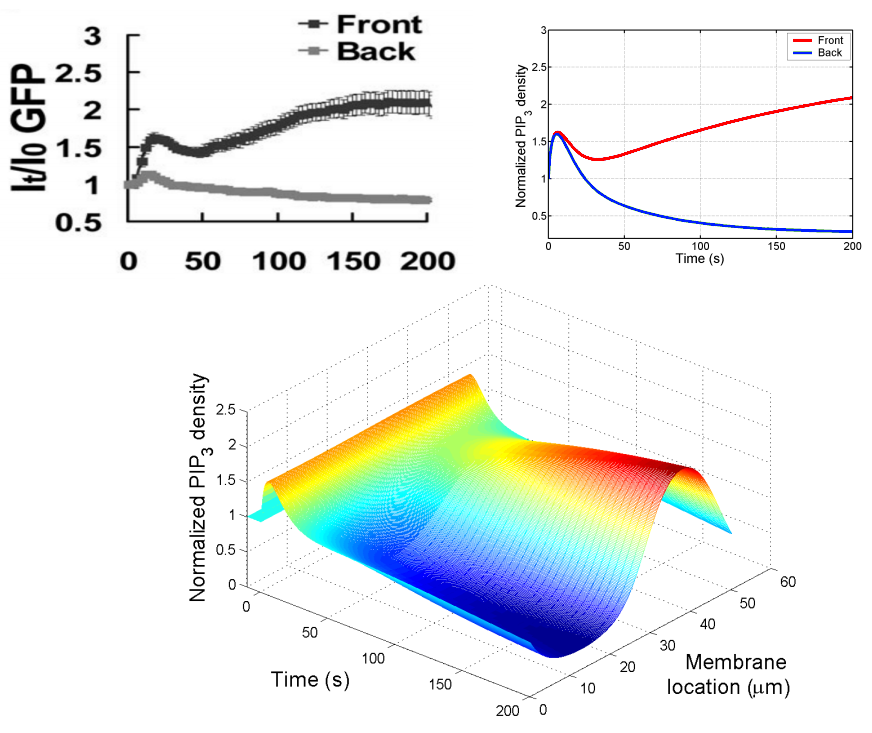}
\caption[\PIPtr\ responses to a static cAMP gradient.]{{\bf $\mathbf{PIP_3}$ responses to a
static cAMP gradient.} {\it (Upper left)} The response at the front and the back
of a cell to a cAMP gradient created by a micropipette. The cell is treated with
latA and assumes a circular shape while the responses are measured by the local
PH$_\tn{Crac}$-GFP concentration \cite{Xu_2005}. {\it (Upper right)} The
simulated \PIPtr\ response on a circular domain subject to a cAMP gradient with
50 \% front-to-back difference. {\it (Lower)} Dynamics of the \PIPtr\ response
along the cell membrane, starting and ending at a location with the mean cAMP
level. The static gradient is applied to a resting cell at 0 $s$.}
\label{circle_grad} 
\end{figure}

To determine how well the model predicts the response to directional cues, we
compare the predictions with the measured \PIPtr\ activity at the front and the
back of an immobilized latA-treated cell reported in \cite{Xu_2005}. There the
cell was subject to cAMP application from a micropipette and exhibited a
biphasic response as shown in Figure \ref{circle_grad}. 
We simulated the experimental stimulation by a step change in cAMP concentration from a uniform
basal level to a static spatial gradient with a mean level of $100~nM$, which
simplifies the spatio-temporal profile of cAMP experienced by the cell. In
reality cAMP reaches the leading edge faster the trailing edge, as shown in
\cite{Dallon_1998}.  The step stimulation is applied to a circular cell whose
upstream Ras module was described earlier and downstream network comprises the
PI3K--\PIPtr\ amplification module shown in Figure \ref{cartoonPI3K} (Equations (\ref{PI3k8})--(\ref{PI3k15})).

Under a directional cue, the localization of steady-state Ras activity is
determined by the ratio between the local activities of RasGEF and RasGAP on the
membrane. Degradation of active RasGEF and RasGAP, which are activated at the
membrane, is characterized by the characteristic decay length $L_d=\sqrt{D/k}$
defined earlier. Differences in the $L_d$ of active RasGEF and RasGAP leads to
differences in their spatial profiles and results in localization of Ras
activity.  In particular, when $L_{d,RasGAP}>L_{d,RasGEF}$, RasGAP is more
evenly distributed than RasGEF in the cytosol, with its level lower than that of
RasGEF at the front and higher at the rear. This establishes a steady-state
gradient of Ras activity that follows the directional cue. On the other hand, if
$L_{d,RasGAP}<L_{d,RasGEF}$ the gradient is reversed.  If the diffusion of
RasGAP is large compared with that of RasGEF then RasGAP is essentially
spatially uniform, and in this case the ratio between Ras activity at the
anterior and the posterior, normalized by the ratio in cAMP levels approaches 1.

The PI3K--\PIPtr\ amplification module combines positive feedback via membrane
recruitment of PI3K and cooperativity due to separate PI3K recruitment and
activation steps, and is capable of significantly amplifying the weakly
localized Ras activity, which is shown in \cite{Khamviwath:2013:MAS}. Parameters
for the PI3K--\PIPtr\ subnetwork are obtained by matching the dynamics of \PIPtr\
response at the front and the back of the cell with an observation in
\cite{Xu_2005}, where a $20~\%$ cAMP gradient across the cell diameter induces
$\tildel180\%$ change in the \PIPtr\ gradient. Figure \ref{circle_grad} displays
the \PIPtr\ dynamics around the membrane of a cell that experiences
$\tildel50\%$ front-to-back difference in cAMP concentration, and compares the
numerical simulation to the experimentally-observed \PIPtr\ activity of a cell
subject to cAMP released from a micropipette.

The response displays the distinctive biphasic behavior characteristic of a
cringe, which occurs in response to directional cues as well as uniform
stimuli. Upon stimulation, \PIPtr\ density over the entire cell membrane
increases rapidly, reaching the first peak within $10~s$. Then the \PIPtr\
density drops throughout the membrane before it starts to rise selectively at
membrane locations above the mean of the cAMP gradient after $\tildel50~s$,
which establishes cell orientation. The first peak is due to the transient Ras
activity that adapts to the mean cAMP level, as the \PIPtr\ response does not
exhibit the transient peak in response to stimulation with a constant mean cAMP
level. On the other hand, the spatial sensitivity displayed by steady-state
\PIPtr\ localization at the leading and trailing edges is due to significant
amplification of small differences in local Ras activity by the PI3K--\PIPtr\
subnetwork. This process is slower than activation and adaptation to the new
cAMP level, thereby producing a distinctive drop in \PIPtr\ level before the
separation between the front and the back sets in. Notice that the observed first
peak at the trailing edge is lower than the peak at the leading edge while the
simulated response exhibits almost identical peaks. This could be due to delayed
exposure to cAMP experienced by the trailing edge \cite{Dallon_1998}, which was
omitted in the simulation.

\begin{figure}[h]
\centering
 \includegraphics[width=4 in]{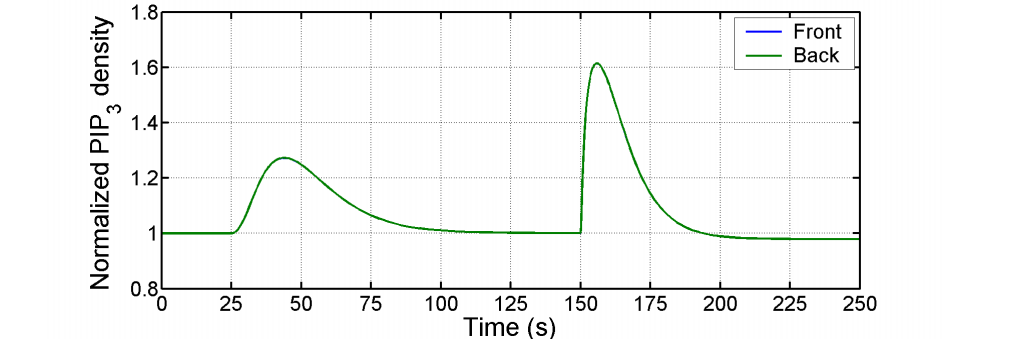}
\caption[Ras-activation dynamics to uniform stimulation at different cAMP
levels.]{{\bf $\mathbf{PIP_3}$ response to repeated stimulation.} \PIPtr\ localization at two
opposite points on the membrane of a circular cell subjected to successive steps
of spatially uniform stimulation. The stimulation is represented by the local
level of free $\Gbg$ which reflects 0.1 $nM$ and 100 $nM$ cAMP at 25 $s$ and 150
$s$ respectively. The responses at the two sites are nearly identical.  }
\label{UnifRepeat}
\end{figure}

Uniform stimulation with repeated increases in cAMP concentration produces
adapting \PIPtr\ responses that are identical in the front and the back of
the cell, as shown in Figure \ref{UnifRepeat}.  
The combined adaptation-amplification mechanism allows consistent localization
of \PIPtr\ activity over several orders of magnitude in mean cAMP
concentration. Thus this direction-sensing model can be used to study
reorientation, polarization, and the roles of PI3K in motility and spontaneous
activity in Dd cells. Note that in our simplified model shown in Figure
\ref{cartoonPI3K} the experimentally-observed positive feedback from F-actin to
Ras activation has been omitted, allowing independent analysis of the effects of
`upstream' Ras activation and `downstream' \PIPtr\ activity. One can include  positive
feedback to Ras so that its activity is sensitive to spatial gradients, while
maintaining adaptation to the mean cAMP level,  and this will be discussed
later.

\begin{figure}[h]
\centering
\includegraphics[width=6 in]{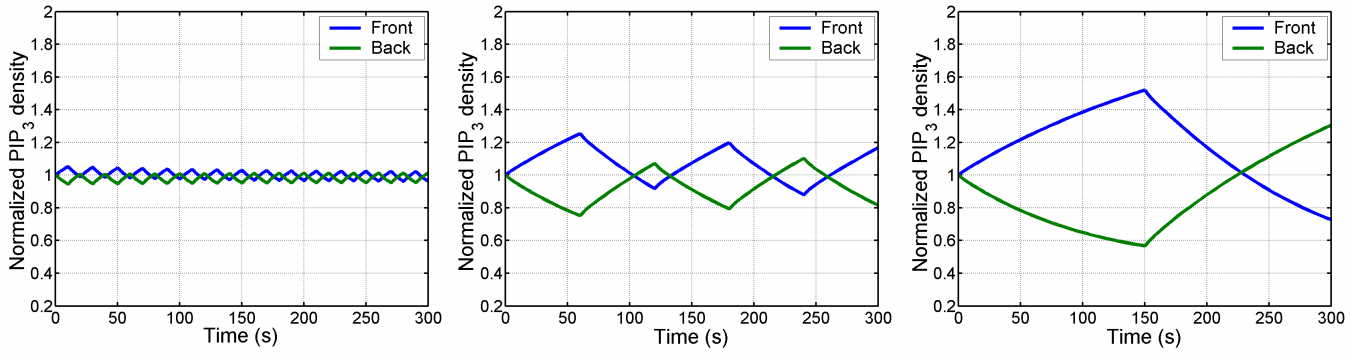}
 \caption[\PIPtr\ responses to alternating gradients at different
periods.]{{\bf $\mathbf{PIP_3}$ responses to alternating gradients of different periods.} {\it
(Left)} 5 $s$, {\it (middle)} 120 $s$, and {\it (right)} 300
$s$.}
\label{circle_alternate} 
\end{figure}

It is known that Dd cells are able to reorient themselves when a cAMP gradient
is reversed. Meier et al. \citet{Meier_2011} observed that the ability to
reorient is limited by polarization dynamics, and \dd\ cells become trapped
under stimulus gradients with rapidly changing direction. Chemotaxing cells move
up the cAMP gradient with reduced speed when subjected to alternating cAMP
gradients with a period of $120~s$ compared to $600~s$. They are completely
stalled and trapped within the alternating gradient when the period is
$20~s$. Figure \ref{circle_alternate} displays simulations of \PIPtr\ dynamics
under alternating gradients at 20 \% difference with periods of 20, 120, and 300
seconds, which agree well with the experimental observations. 
The response develops a very small front-to-back gradient at the highest frequency, which
explains its inability to polarize under the rapidly-alternating gradient. As
the frequency decreases, stronger \PIPtr\ localization gradients are allowed,
leading to the experimentally-observed increase in the chemotaxis speed.

\subsection*{The roles of cellular shapes in polarization and motility}

Polarized cells are elongated and have a well-defined anterior and posterior. In
the absence of directional stimulation, they move persistently in the direction
of their anterior and only change direction significantly every 9 minutes on average. They migrate by
alternately splitting left and right pseudopods from their leading edge, between 
$40^\circ$ and $70^\circ$ to their polarization axis, and occasionally develop {\it de novo}
pseudopods which cause abrupt changes in the movement direction
\cite{Andrew_2007, Li_2008, Bosgraaf_2009}.  Experimental observations suggest
that the polarization of Dd and neutrophils depends on PI3K activity and that
these cells become more polarized at higher levels of persistent uniform
stimulation \cite{Weiner_2002, Postma_2004, Haastert_2004, Ferguson_2007}.  To
study the PI3K activity in polarized cells, we apply various types of cAMP
stimulation to 2D simulation domains which resemble the shape of polarized cells
on the substrate surface. Since shape changes are slower than biochemical
re-polarization, we study the chemotactic responses in frozen domains as a first
step to approximate the dynamics of signaling proteins in motile cells. Recent
work \cite{Khamviwath_2013} on the network downstream of \PIPtr\ that leads to
F-actin polymerization will be discussed later.

\begin{figure}[h]
\centering
 \includegraphics[width=6 in]{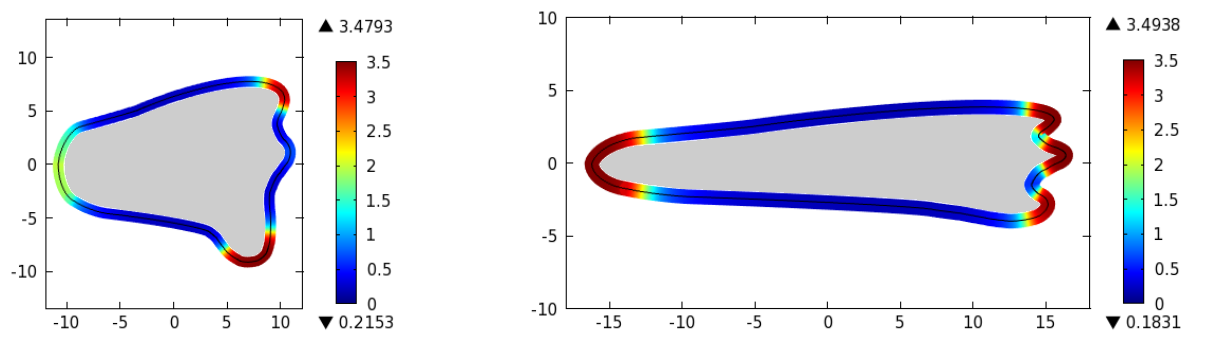}
\caption{{\bf The steady-state $\mathbf{PIP_3}$ localization at the boundary of cells subject to uniform
stimulation at 100 nM cAMP.} {\it (Left)} A moderately polarized cell. {\it
(Right)} A highly polarized cell. Reaction-diffusion equations are solved in the regions shown. 
Color bars display \PIPtr\ activity at the boundary, normalized by its unstimulated level. 
Boundary curve is inflated to show color variations.}
\label{geom_unif}
\end{figure}

\begin{figure}[h]
\centering
 \includegraphics[width=6 in]{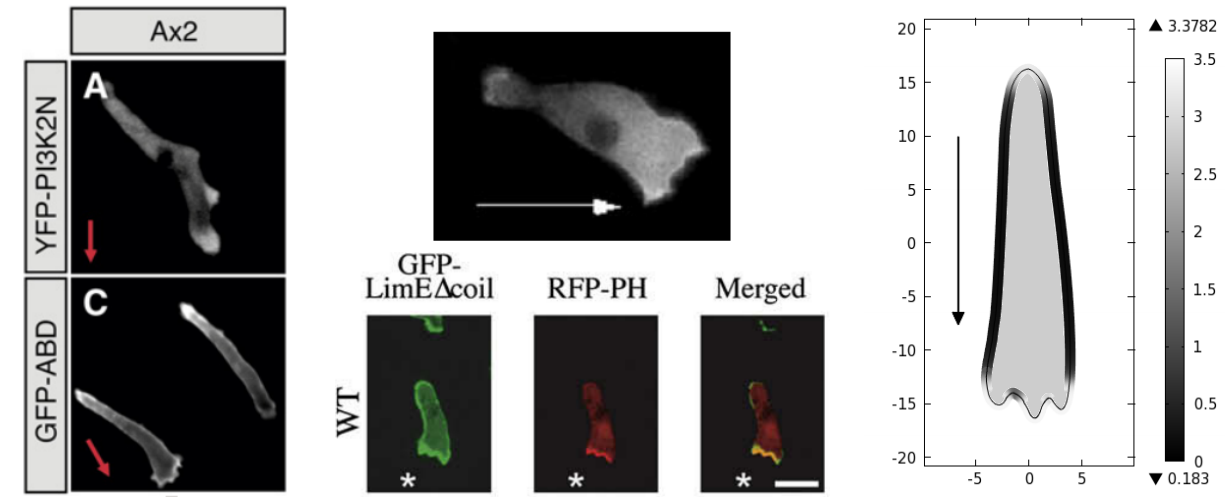}
 \caption{{\bf Localization of signaling molecules in highly polarized cells
migrating towards cAMP sources.} The directions of cAMP gradients are indicated
by arrows. Stars indicate tips of micropipettes. {\it (Left panel)} PI3K
localization {\it (top)} and F-actin activity {\it (bottom)} are the highest at
the posterior and at protrusions and membrane ruffles near the anterior
\cite{Hoeller_2007}. {\it (Middle panel, top)} RacB activity, which is
downstream of \PIPtr\, is high at the front and noticeable at the back of the
cell \cite{Park_2004}. {\it (Middle panel, bottom)} F-actin {\it (green)} and
\PIPtr\ {\it (red)} activities are  highest near the micropipette. The
activity at the back of the cell is also above the normal level
\cite{Zhang_2008}. {\it (Right panel)} Simulated \PIPtr\ activity under a 50 \%
front-to-back gradient is the highest at the anterior. There is also significant
\PIPtr\ activity at the posterior.}  
\label{geom_directional} 
\end{figure}

Our simulation results suggest that biochemical polarization is at least partly
determined by the spatial  configuration of the cell, particularly the
curvature of the membrane.  Figure \ref{geom_unif} depicts the steady-state
\PIPtr\ localization of polarized cells subject to uniform stimulation.
\PIPtr\ is localized strongly at the left and right protrusions at the leading
edge of the moderately polarized cell, with stronger \PIPtr\ localization in the
right pseudopod, in good agreement with observations which show that pseudopod
extension occurs at the left and the right of the leading edge
\cite{Bosgraaf_2009}. On the other hand, \PIPtr\ is strongly localized at all
anterior protrusions in the strongly polarized cell, suggesting a more unimodal
distribution in the direction of pseudopod extension.  Interestingly, \PIPtr\ is
also localized, moderately and strongly, at the trailing edge of the moderately
and strongly polarized cells, respectively. Simulated \PIPtr\ localization in
cells polarized in the direction of increasing cAMP is consistent with observed
localization of many signaling molecules within the pathway, including F-actin,
PI3K, \PIPtr, and RacB, in cells migrating towards directional cues
\cite{Zhang_2008, Hoeller_2007, Park_2004}. Figure \ref{geom_directional}
compares the simulated PIP$_3$ localization to the experimental observations.

\begin{figure}[h]
 \centering
\includegraphics[width=6 in]{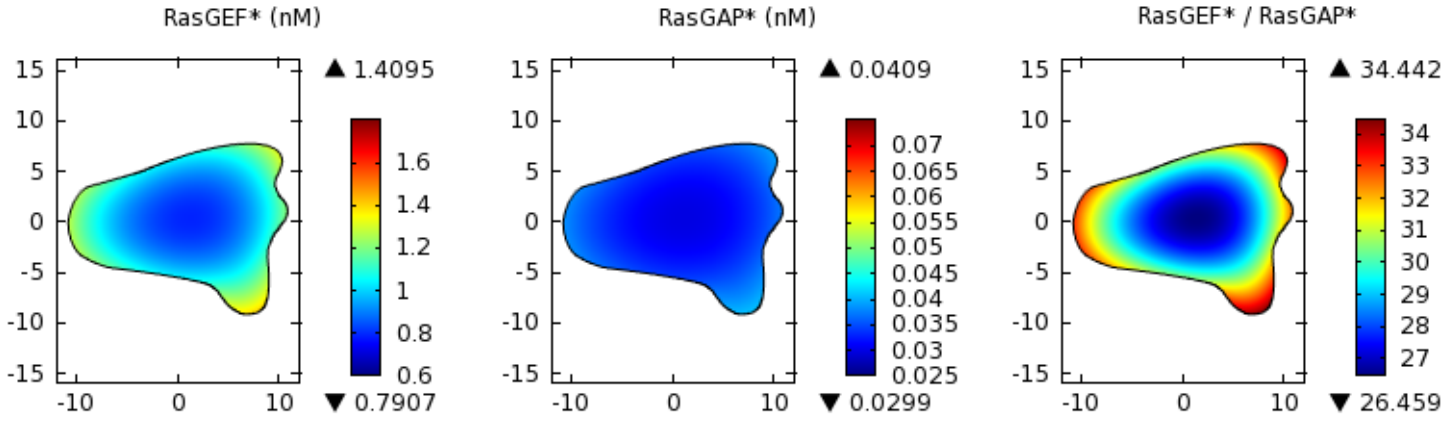}\\
\caption{{\bf The steady-state activities of RasGEF and RasGAP under uniform stimulation.}}
\label{gefgap} 
\end{figure}

\begin{figure}[h]
 \centering
\includegraphics[width=6 in]{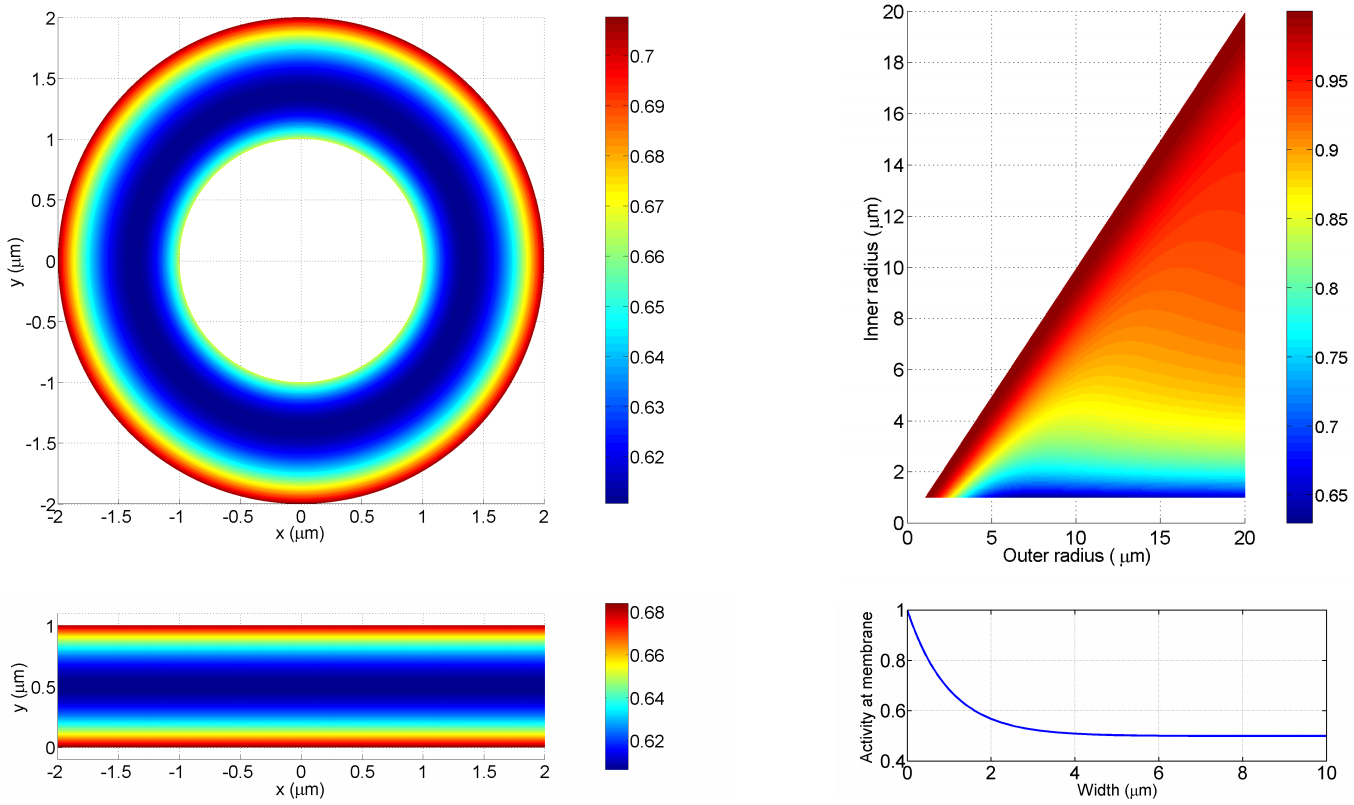}\\
\caption{{\bf The steady-state activity of a membrane-activated protein in  a cross
 section through the center of a 3D shell \textit{(upper)} and an infinite strip
 \textit{(lower)}.} {\it (Right column, upper)} The ratio between the protein
 activity at the inner and outer radii of shells with various sizes. {\it (Right
 column, lower)} The protein activity at the boundary of infinite strips of various
 width. In these simulations, $L_a=L_d=1~\mu m$, where the characteristic
 lengths $L$ are defined in the Methods section.}
\label{cartoonAct} 
\end{figure}

Simulated \PIPtr\ activity influenced by the cell shape tends to localize in 
re-entrant regions of the boundary, such as tips or narrow tethers, in
good agreement with measurement of \PIPtr\ localization in fibroblasts
\cite{Schneider_2005}. Examination of Ras activity, which exhibits subtle
localization under uniform stimulation, suggests that the pronounced \PIPtr\ 
localization is due to amplification of the active Ras spatial profile. 
Figure \ref{gefgap} displays the spatial localization of RasGEF and RasGAP activities and
 their ratio at the steady state. 
Both activated RasGEF and activated RasGAP are
 localized in re-entrant regions, but activated RasGAP is more uniformly
 distributed. Their ratio at the boundary, which determines the Ras activity at
 the membrane, is also higher in these regions than elsewhere.
To rationalize the \PIPtr\
localization caused by Ras activity, which is in turn regulated by active RasGEF
and RasGAP, we sought to understand how the cell shape affects localization of
RasGEF and RasGAP activities under uniform stimulation. We analyzed the
steady-state activation profiles of a membrane-activated molecule in a 3D shell
and a thin strip to better understand the effect of the mean curvature of the domain and the
distance between boundaries, respectively, on localization.  These cartoon descriptions allow
for analytical solutions (see the Methods section) which are plotted in Figure
\ref{cartoonAct}.

In the case of the 3D shell, the difference in curvature contributes to the
difference in the activity at the inner and outer membrane, where the convex
(viewed from the interior of the shell) outer membrane is less exposed to the bulk cytosol
than the concave inner membrane, leading to higher activity at the outer
membrane. The ratio between the outer and inner activity depends on the
difference in curvature and the distance between the inner and outer
membrane. For an infinite strip, the membrane activity decreases as the strip
becomes wider because the activity at one boundary has less effect on the other
side. It approaches a constant when a molecule activated at one side fails to
reach the other side before it becomes inactive. If we subtract the asymptotic
(in the width) amplitude from the result in the lower right, and fit the
remainder with an exponential, the decay constant in membrane activity as a
function of the domain width approximately coincides with $L_d$. In summary, we
found that membrane along a thin region may have much higher activity than
membrane within the same cell that is well exposed to cytosol, and that membrane
curvature contributes to its cytosolic exposure. Moreover, an exposed membrane
region may also have high activity if it is in close proximity to a region with
high activity. Ras and \PIPtr\ localization follows the localization of active
RasGEF and RasGAP, since $L_{d,RasGAP}>L_{d,RasGEF}$, and small variations in
RasGEF and RasGAP activity may ultimately lead to highly-polarized \PIPtr\
activity due to amplification by the PI3K--\PIPtr\ subnetwork. Thus when a cell
is subjected to a directional cue, the spatial distributions of active RasGEF
and RasGAP are dictated by both cell shape and stimulation, in effect
integrating the intrinsic polarity with the external information to produce `biased'
\PIPtr\ polarization. This intrinsic polarity that derives from the cell shape
is in contrast to explicit polarity studied in \cite{Krishnan_2007, Hecht_2011}.

\begin{figure}[h]
\centering
 \includegraphics[width=5 in]{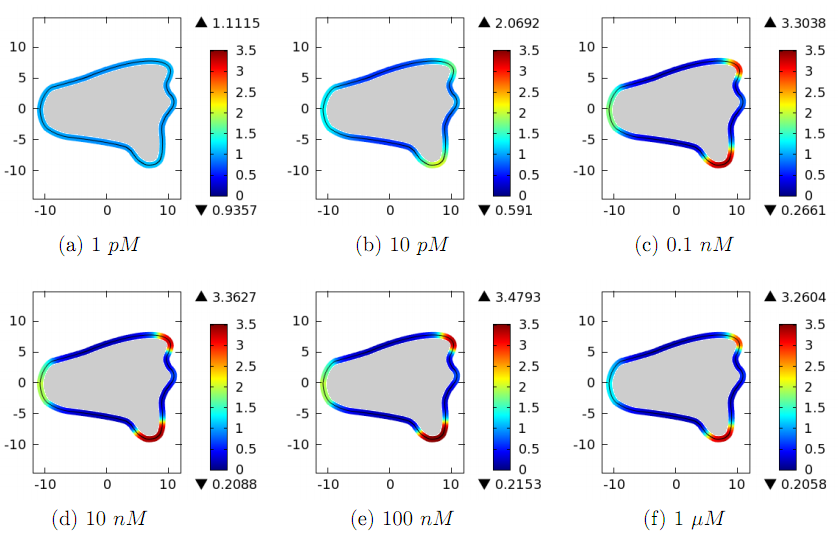}
\caption[Dose dependence of steady-state \PIPtr\ localization in a polarized
cell.]{{\bf The steady-state $\mathbf{PIP_3}$ localization in a moderately polarized cell at
different levels of uniform stimulation.}} 
\label{geomspon_unif}
\end{figure}

\begin{figure}[h]
 \centering 
\includegraphics[width=5 in]{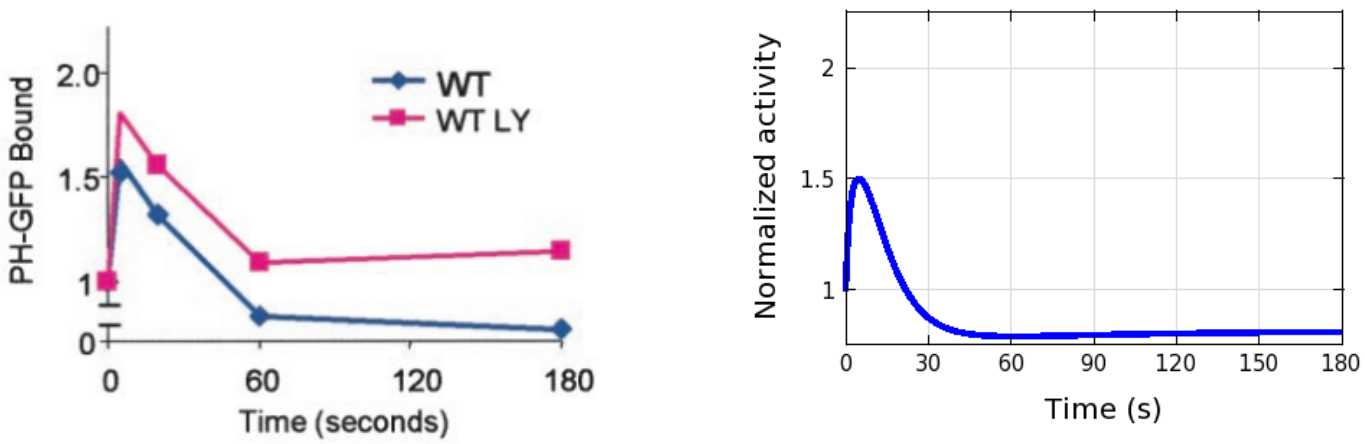}
\caption[Dynamics of overall \PIPtr\ responses in wild-type cells.]{{\bf Dynamics of
overall $\mathbf{PIP_3}$ responses in wild-type cells. Uniform stimulation with 1 $\mathbf{\mu M}$
cAMP is applied at 0 $\mathbf{s}$.} {\it (Left)} Experimental measurement in wild-type
(WT) and LY294002-treated cells \cite{Chen_2003}. {\it (Right)} Simulation of
the moderately polarized cell.}
\label{OverallActin} 
\end{figure}

The shape-induced polarization of \PIPtr\ is dependent on the mean cAMP level,
as shown in Figure \ref{geomspon_unif} for spatially-uniform stimulation at
various cAMP levels. The dependence arise from the balance between 
spontaneous GTP hydrolysis and turnover of Ras and cAMP-dependent
RasGEF and RasGAP activities, which are affected by cell shape.
 One can see that the degree of polarization is an increasing function of the
stimulus level except at 1 $\mu M$, where RasGEF begins to saturate ({\it cf.} Figure \ref{RasAdapt}). 
Wang et al.\cite{Wang_2012} observed that removal of the stimulus leads to
a transient decrease in the \PIPtr\ level, which is consistent with the recovery
time of approximately 3 minutes that is required before fully-adapted Dd cells become
responsive to new stimulation at the previous cAMP level \cite{Dinauer_1980,
Xu_2007, Wang_2012}. While a simple feedforward control leads to  slow recovery
after cAMP removal, which depends on the cAMP level, the modulation between spontaneous
and cAMP-induced regulation of Ras ensures the recovery period for the signaling
pathway is consistent with {\it in vivo} observations.  Note that although
\PIPtr\ activity is high at many locations along the cell membrane, total
\PIPtr\ activity may drop slightly compared with the pre-stimulation level
because saturation causes lower average activity within the cell, as observed
{\it in vivo} \cite{Chen_2003}. Figure \ref{OverallActin} displays average
\PIPtr\ dynamics within $180~s$ after uniform stimulation.

In the absence of directional cues, polarized cells migrate persistently in the
direction of their polarity. Our model suggests that there is positive feedback
between cell polarity encoded by the cell shape and intracellular signaling that
leads to F-actin localization and promotes pseudopod extension. Small
protrusions at the anterior induces localization of \PIPtr\ and F-actin, which
in turn drive the extension, contributing to directional persistence. Moreover,
in moderately polarized cells, \PIPtr\ tends to localize within the protrusion
at the left and right ends of the anterior, where membrane length per cytosolic
area is the highest. This selective localization leads to high propensity of
extending new pseudopods at an angle to the polarization axis. If there is a
slow negative feedback to the F-actin localization process, due for example to scarcity
of Arp2/3 and G-actin, as described in \cite{Machacek_2006, Khamviwath_2013} that
causes retraction of older pseudpods, the overall process could result in the
characteristic zig-zag movement observed {\it in vivo} \cite{Andrew_2007,
Li_2008, Bosgraaf_2009}.  

\begin{figure}[h]
\centering
\includegraphics[width=5 in]{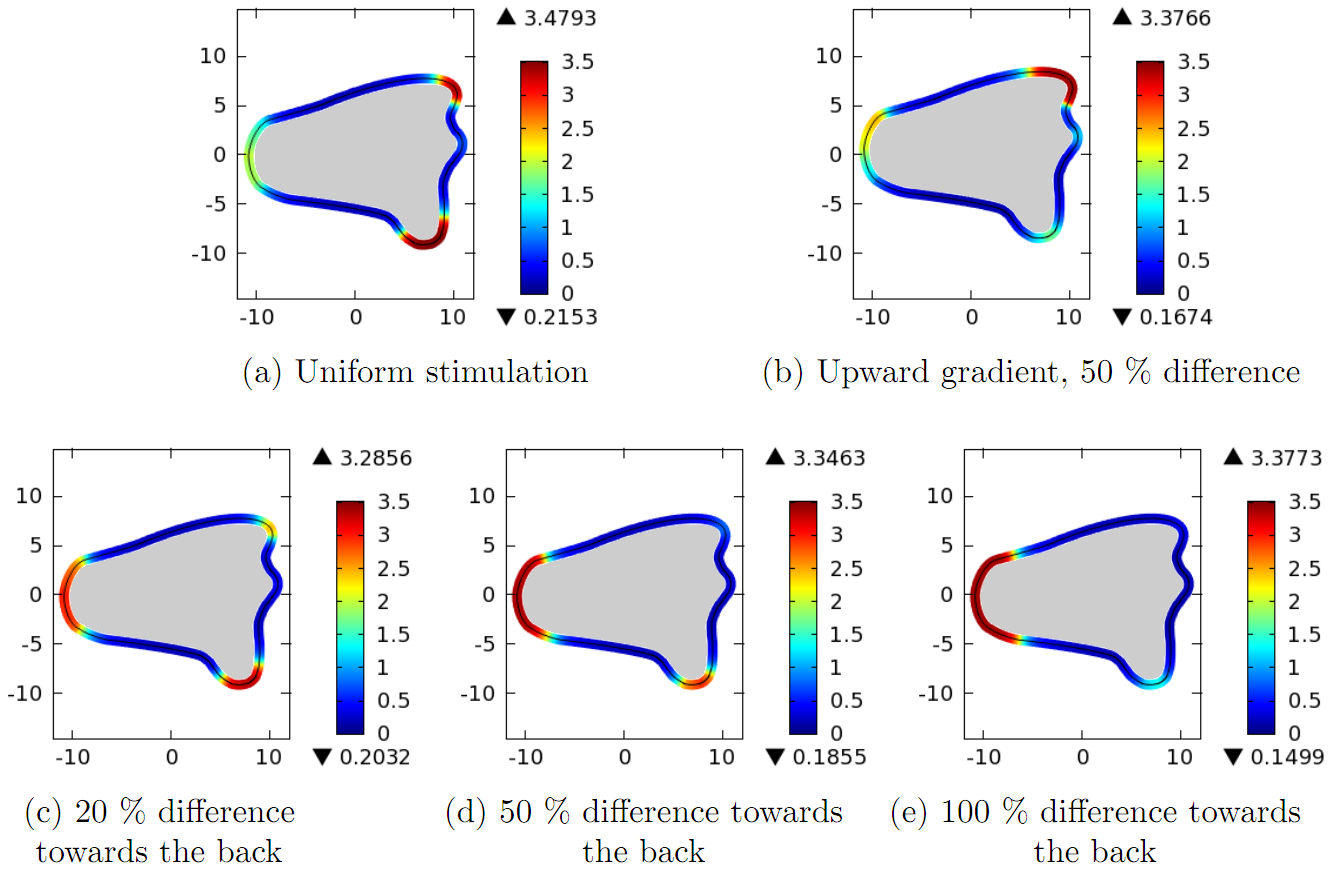}
 \caption{{\bf The steady-state $\mathbf{PIP_3}$ localization of a polarized cell induced by
  static cAMP gradients.} \figlabel a A polarized cell subjected to uniform
  stimulation displays highest \PIPtr\ localization at the lower
  pseudopod. \figlabel b A 50\% cAMP gradient across the cell, along the \textit{y}-axis, biases the
  localization towards the upper pseudopod. \figlabel{Lower row} \PIPtr\
  localization subject to backward cAMP gradient, along the negative \textit{x}-axis, at different 
  levels suggests distinct modes of movement. \figlabel c biased movement induced by
  20\% gradient across the cell. \figlabel d and \figlabel e The cell reorients
  its front under large cAMP gradients.  }
\label{geom_grad_back} 
\end{figure}

When subject to a directional cue, RasGEF and RasGAP
activity is determined by both the cAMP gradient and shape-induced
polarity. Unlike a rounded latA-treated cell, which reorients its \PIPtr\
localization directly upwards a cAMP gradient, a polarized cell exhibits \PIPtr\
localization that is influenced by both factors. Usually, polarized cells
maintain their polarity and make gradual turns towards the cAMP source and
reorient themselves only when they are subjected to a strong cAMP gradient
\cite{Andrew_2007, Haastert_2004, Chen_2003}. We can determine the chemotactic
responses in polarized cells predicted by the model by applying cAMP gradients
at different levels and directions. Figure \ref{geom_grad_back} displays the 
steady-state \PIPtr\ localization biased by the gradients.
Under uniform stimulation, the \PIPtr\ localization is the highest at the right
pesudopod. 
Figure S1 depicts the localization dynamics. 
A $50\%$-gradient towards the top induces a clear directional bias
for the left pseudopod. A 20\% backward gradient cannot overcome intrinsic
polarity and the \PIPtr\ localization at the right pseudopod remains the
strongest. At steeper gradients, \PIPtr\ localization directly orients towards
the back of the cell. Taken together, the numerical results suggest that a directional gradient
normal to the polarity axis induces a turn towards the cAMP source by extending
a pseudopod from the anterior at the position closest to the source. The cell
gradually turns around maintaining its anterior and posterior under a shallow
gradient that directs backwards while it reorganizes a new anterior directly at
the back when subject to a strong backward gradient. Figure S2 shows PIP$_3$
localization dynamics corresponding to Figure \ref{geom_grad_back}b. In the
first phase, PIP$_3$ localization peaks throughout the membrane due to
adaptation to the mean cAMP level. Then the PIP$_3$ level drops and selectively
localizes at the top of the leading edge, implying the cell will continue
migrating forward, with an upward bias.  

An interesting experiment by Houk et al. \cite{Houk_2012} suggests that membrane
tension, and not protein regulation, plays the role of global inhibitor for
local protrusion. In this experiment, cells were subject to a heat shock,
causing them to assume an irregular shape where the front and the back were
separated by a long and narrow tether. Under the hypotheses of the biochemical model, activator and
inhibitor would have been trapped in the front and the back respectively, the
back should not have been able to resume spontaneous protrusion. However, after
the tether is severed, the former back resumes spontaneous pseudopod extension,
suggesting that the inhibition is not biochemical. We investigate if our signal
transduction model is able to explain this interesting behavior. In our
simulations, less than 20\% of PI3K and PTEN is bound to the membrane while
total concentration of RasGEF and RasGAP is uniform in the cytosol. Therefore,
by separating the front and the back, only the concentration of PI3K and PTEN will 
change. According to our model, the change in PI3K and PTEN density alter
the sensitivity curve for \PIPtr\ localization. The former front will generally
have more \PIPtr\ activity while the former back will have less \PIPtr\
activity. Nevertheless, both halves are still capable of forming spontaneous
protrusions. In fact, this agrees well with the observations in \cite{Houk_2012}
as the former front undergoes excessive protrusions after it becomes free. Note
that while the cell assumes the tethered shape, membrane tension could indeed
play a major role in restricting protrusions within the tether and the back.

\section*{Discussion}

We developed a modular model for the network involved in signal transduction and
the first steps in control of the actin network in eukaryotic cells. The model
incorporates biochemical interactions that are well-established in \dd\ and captures many
aspects of its responses to cAMP stimulation. The model consists of
adaptation and amplification modules that  are responsible for regulation of
Ras and \PIPtr\ activities, respectively.  Simulations of this model give
insights into dependence of cell polarization on mean stimulus levels, how it is 
embedded in cell shapes, how it influences
pseudopod extension and creates zig-zag movement pattern, and how it integrates with
directional signals and gives directional persistence.

The adaptation module leads to rapid excitation and slower adaptation of Ras
activity by feedforward regulation of its activator (RasGEF) and its inhibitor (RasGAP)
over the relevant range of cAMP stimuli. Although it is unclear how RasGAP is
activated by the external stimulus, the feedforward regulation is a simple
scheme which serves well for adaptation of a molecular switch like Ras. Ras
activity is able to adapt to repeated uniform stimulation, and the extent of
adaptability is determined by saturation of either RasGEF or RasGAP activity.
Recently-observed over-adaptation of Ras activity at high cAMP levels indicates
that the inhibition signal is not downstream of Ras, and supports a control
scheme such as we use, in which RasGEF activity is saturated before RasGAP
activity \cite{Takeda_2012}. Over-adaptation serves a useful purpose in that it
serves  to prevent elevated F-actin activity at saturating cAMP levels. When a
stimulus is removed, there is an under-shoot of the Ras activity
\cite{Xiong_2010}, and a recovery period is required before the cell becomes fully active
to the previous stimulation level. This recovery time is dependent on the
spontaneous GTP hydrolysis activity of Ras.

We found that the observed transient peak in F-actin shortly after the stimulus
is applied is largely due to the adaptation mechanism, while the second,
less-pronounced, phase of the chemotactic activity is due to amplification of
small spatial variations of Ras activity along the membrane.  In the feedforward
model, the transient peak in Ras activity is caused by fast activation of RasGEF
and slow activation of RasGAP.  In the presence of a cAMP gradient, the
assumption that  RasGAP has a longer decay length that RasGEF leads to  a stable 
gradient in Ras activity that mirrors the cAMP gradient. 
Our model differs from the activator-inhibitor type in that we do not assume
that RasGEF diffuses much more rapidly than RasGAP -- in fact they have equal
diffusion coefficients in our simulations. 


The amplification module involves the regulation of \PIPtr\ by PI3K and PTEN,
and positive feedback that is sensitive to latA (which suggests dependence on
F-actin). The activity of PI3K depends on both its membrane localization due to
the F-actin activity and subsequent activation by Ras \cite{Funamoto_2001}. This
two-stage activation of PI3K is sufficient to induce a greatly-amplified spatial
gradient of \PIPtr\ activity without the usual assumptions of cooperative
binding. Analysis shows that this network structure allows a sigmoidal response
with arbitrarily high amplification \cite{Khamviwath:2013:MAS}. Therefore small
variations in Ras activity can be amplified into significant \PIPtr\ gradients
along the membrane, thereby ensuring a suitable directional response. The
biphasic response is a result of distinct time scales for adaptation and the
\PIPtr\ dynamics, which becomes apparent after $\tildel50~s$ and takes several
minutes to  fully develop. The time scale for developing \PIPtr\ orientation
coincides with the ability of \dd\ to reorient and migrate under rapidly
alternating cAMP gradients, suggesting an important role for  \PIPtr\ activity in
cell orientation \cite{Meier_2011}.

In this work we incorporated only the structure necessary to produce the
observed responses, and have omitted positive feedback that affects Ras activity
and causes localization of RBD to spontaneously-extended pseudopodia
\cite{Sasaki_2007}. In fact,  this second positive feedback loop (see Figure
\ref{network}) is also needed to produce significant Ras localization under
directional stimulation.  If Ras activation via this positive feedback
loop acts independently of activation by the external stimulus, for example by
independent sites on Ras,  adaptation will be preserved under suitable conditions 
on the feedback. In this case, any activity of \PIPtr\ and
F-actin that occurs spontaneously will be observed in Ras as well. An analysis
of adaptation with positive feedback from F-actin under the foregoing condition
is given in the Methods section, and more detailed descriptions of the pathway 
are the subject of future work.
Although this work is focused on the Ras--PI3K 
interaction (see Figure \ref{cartoonPI3K}), there may be other inputs for the 
PI3K pathway such as Rac activation via ElmoE. Each input may trigger the feedback 
network shown in Figure \ref{network} and cause directed migration. For example, 
photoactivation of Rac can direct neutrophil migration and it was demonstrated that 
PI3K is critical to this activity. \cite{Yoo_2010}

We have shown that asymmetric cell shapes can lead to spatially non-uniform Ras
activity, which is then amplified into \PIPtr\ gradients. In this case, \PIPtr\
localization is determined by a combination of chemotactic and geometric
factors, the latter of which can be considered an intrinsic polarization of the
cells. 
In narrow regions of a cell, activated RasGEF is more localized
near the boundary than activated RasGAP, causing higher than average local Ras activity.
This integration of cell polarity with
localization of \PIPtr\ and its effectors suggests a role for membrane ruffles
within the cell anterior in maintaining polarity of the cells. Similarly,
filopodia may serve as precursors for pseudopodia. Moreover, because \PIPtr\
tends to localize more strongly in regions where the membrane curvature is
large, this may cause polarized cells to preferentially extend pseudopodia at a
large angle to the polarization axis if small regions of high curvature develop
there. If these extensions in a hypothetical cell are out of phase, the cell
could alternately extend pseudopod branches and display the zig-zag migration
pattern observed in \dd\ chemokinesis \cite{Li_2008, Bosgraaf_2009}. In
contrast, more polarized cells without ruffles or other local small protrusions
have \PIPtr\ concentrated throughout the leading edge, and will migrate 
more directly toward the stimulus. Because \PIPtr\ localization amplifies the Ras 
distribution, which is a combination of intrinsic polarity and external stimulation,
the balance between them determines modes of directed migration. In a shallow
cAMP gradient, polarized cells  turn towards the stimulus source slowly, while
maintaining their leading edge. On the other hand, when the directional cue
overcomes the bias, \PIPtr\ is most strongly oriented towards the source,
causing the cells to develop new anterior in this direction.

Neutrophils and starved \dd\ cells become more polarized and have increased
motility under uniform stimulation \cite{Zigmond_1977, Kriebel_2003,
Postma_2004}. These behaviors are dependent on PI3K activity
\cite{Funamoto_2001, Ferguson_2007, Afonso_2011}, and our model suggests that
such activity can be a result of modulation between small spontaneous self-activation and
actively-regulated Ras activity, and that the polarity of the \PIPtr\ signal is
dependent on the cell conformation, creating a positive feedback between
biochemical and physical attributes of the cell. The implications of this are
two-fold. Firstly, this positive feedback could potentially lead to spontaneous
pseudopodia extension. The onset of spontaneous protrusions could be the 
result of F-actin waves, which perturb membrane
conformation.  In fact, observations in \dd, neutrophils, and fibroblasts suggest
that pseudopod extension frequency is dependent on the stimulation through the
PI3K pathway and that breaking the feedback loop by PI3K inhibition disrupts the
ability to sustain polarity, even when local protrusions are induced by
photo-activation of Rac, which is downstream of \PIPtr\ \cite{Welf_2012,
Yoo_2010, Postma_2004}.  Secondly, the fact that polarization of \dd\ is dependent
on the external cAMP level may link polarity to signal relay in starved \dd\
cells, where secretion of cAMP serves as a means to develop self-induced
polarity \cite{Kriebel_2003}.

In this work, we elucidated how the cell shape and intracellular signaling are 
related to cell motility and polarization. In future work, we will
incorporate cell movement into the current model to better understand the
interplay between signaling and the cell shape and how this determines extension
and survival of pseudopods. Spontaneous actin waves associated with PI3K
activity as well as spontaneous \PIPtr--PTEN dynamics have been observed in \dd\
and could play roles in driving spontaneous pseudopod extensions and retractions
\cite{Gerisch_2010, Arai_2010}. Our signaling model incorporates neither the
positive feedback through Ras nor the cooperativity within the PI3K feedback
loop due to branching of F-actin and is not excitable, despite exhibiting high
gradient amplification. We have studied the actin waves using a model which
incorporates F-actin branching \cite{ Khamviwath_2013} while PI3K-based models
which include cooperativity have been proposed to study the \PIPtr--PTEN dynamics
\cite{Arai_2010, Taniguchi_2013}. However, external stimulation has not been
incorporated in these models. A more comprehensive picture of signaling dynamics
regulating cell polarization and movement will likely  integration of all
these aspects of the system.

Because of its simplicity, our model cannot account for some  aspects of the
chemotactic responses. 
It is known that latA treatment leads to suspended spherical cells that have minimal F-actin 
activity. When we substantially reduce the
positive feedback from \PIPtr\ to PI3K localization, the spatial sensitivity is
diminished, but is recovered when we adjust system parameters so
that the system relies on the positive feedback between PTEN and PIP$_2$. It may be possible
to obtain a parameter set which fully utilizes both feedback loops so
that the system remains highly sensitive to spatial gradients even when the
F-actin activity is severely reduced. Furthermore, the F-actin branching process
possesses intrinsic cooperativity and positive feedback which can provide
additional amplification to the PI3K--\PIPtr\ feedback loop. An additional
positive feedback from F-actin activity to Ras activation has been observed and
can contribute to the sensitivity of the PI3K pathway. Next, after cAMP removal,
\dd\ cells become insensitive to stimulation up to the previous level for
several minutes. This is likely due to a negative feedback from the downstream
circuit to the adaptation module. This negative feedback could also explain the
transient polarity reversal observed when uniform stimulation is applied shortly
after removal of a static cAMP gradient \citet{Xu_2007}. However, currently
there is no plausible candidate which negatively links the downstream activities
to Ras activation. 

Another aspect of signaling and network dynamics that has not been addressed
here concerns the role of stochastic fluctuations.  We have assumed that the
system is deterministic, but certainly the number of cAMP molecules near a cell
fluctuates, hence the number of bound receptors and downstream components all fluctuate. Estimates
reported in \cite{Othmer:1998:OCS} show that if the number of molecules in a
`capture region' surrounding a cell is in the hundreds, which obtains at low
signal levels, the fluctuations in receptor occupancy will be significant, but at
high signal levels they will not be important. In any case, small numbers of
molecules of components in the downstream pathway may be significant. In fact,
it has been shown that the stochastic version of the actin wave model described
above gives rise to most of the phenomena observed during the re-building of the
actin network following treatment with latA, and this suggests that
stochastic effects may be important in the random extension of pseudopods in the
absence of directed signals. We are however some distance from a stochastic
model that integrates signaling and mechanics.

In summary, many aspects of the chemotaxis responses  can be explained by a simple
model which encompasses known interactions between components of the
pathway. The model also illuminates an important role of the cell morphology 
in affecting how \dd\ cells react to external stimulation. It also suggests a
possible function of membrane ruffles and filopodia at the leading edge of \dd\
cells. Recent studies have linked PLA2 and sGC to directional persistence
\cite{Bosgraaf_2009, Bosgraaf_2009a}, and it is possible that they are involved in
formation of these irregular membrane structures. The dose-dependent polarity
indicated by the model also suggests a new role of \dd\ cAMP secretion in self
polarity enhancement. However, future studies are needed to understand more
subtle behaviors such as the transient inverse polarization and how
intracellular signaling interacts with cell movement. After all, the PI3K
pathway is only one of the parallel pathways that contribute to the overall
chemotactic responses. It remains to be determined how concerted activities of
these pathways lead to chemotactic behaviors of \dd\ cells, and models will play
an important role in understanding this.

\newpage
\section*{Methods}

\subsection*{The model for PIP$_3$ adaptation and gradient amplification}

The biochemical network underlying the model consists of an adaptation
subnetwork and an amplification subnetwork, as shown in Figure
\ref{cartoonPI3K}, whose outputs are Ras activity and PIP$_3$ activity,
respectively.  The input to the model is free $\Gbg$ density on the
membrane. For simplicity, we assume that 50\%, or 1000 $\#/\mu m^2$, of total
$\Gbg$ is free under uniform stimulation at 1 $\mu M$ cAMP and that free $\Gbg$
density is linearly dependent on cAMP concentration
\cite{Meier-Schellersheim_2006}.  Interconversion between different forms of the
same molecule is denoted by solid arrows while positive regulation and promotion
of a particular species and process are denoted by dashed arrows. In Figure
\ref{cartoonPI3K}, membrane-bound species are written in bold while other
species are located in the cytosol. The inactive forms of RasGEF and RasGAP are
part of the model but are omitted from the diagram as their conversion from
active forms into inactive forms is spontaneous. In this model, all spontaneous
activation and inactivation are assumed to be negligible unless these activities
are important to the response. We explicitly include spontaneous activation and
inactivation of Ras, which lead to polarity-induced \PIPtr\ localization, and
spontaneous membrane binding of PI3K, which is crucial for high spatial
sensitivity of the \PIPtr\ response. Note that contribution of these spontaneous 
interconversions are very small compared to that of their regulated activation and inactivation.  In
reality, a regulator forms a complex with its substrate before it may convert or
activate the substrate. We assume for simplicity that all complex formation is
fast and negligible amount of molecules is in the complex form so that the
conversion rate of the substrate is proportional to the product of regulator and
substrate densities. This assumption applies to membrane reactions, cytosolic
reactions, and reactions at the cytosol-membrane interface. This simplification
is discussed in a later section. Since diffusion of cytosolic species is
significantly faster than diffusion of membrane-bound species, we assume no
membrane diffusion.

The numerical simulations of model dynamics are done  on 2D domains
($\Omega$) including a circular disk and more realistic cell shapes (\textit{cf.} 
Figure \ref{geom_unif}). We assume
conservation of signaling molecules and that they are initially uniformly
distributed. This assumption implies that the sums of active and inactive forms
of RasGEF and RasGAP remain uniform throughout the simulations when the active
and inactive forms diffuse equally fast. For each simulation, we apply low basal
level of input, which is equivalent to $0.1~ pM$ cAMP and allow the system to
reach its equilibrium before applying stimulation.  A full set of reactions
which describe our model for PIP$_3$ activity consists of
\begin{align}
\label{Ras1}
\Gbg+RasGEF&\stki{k_{RasGEF^*}} \Gbg+RasGEF^* \quad &\tn{on } \del\Omega\\
RasGEF^*&\stki{k_{RasGEF}} RasGEF \quad &\tn{in } \Omega \\
\Gbg+RasGAP&\stki{k_{RasGAP^*}} \Gbg+RasGAP^* \quad &\tn{on } \del\Omega\\
RasGAP^*&\stki{k_{RasGAP}} RasGAP \quad &\tn{in } \Omega\\
RasGEF^*+Ras&\stki{k_{Ras^*}} RasGEF^*+Ras^* \quad &\tn{on } \del\Omega\\
RasGAP^*+Ras^*&\stki{k_{Ras}} RasGAP^*+Ras \quad &\tn{on } \del\Omega\\
\label{Ras7}
Ras&\stk{k_{s,Ras^*}}{k_{s,Ras}} Ras^* \quad &\tn{on } \del\Omega\\
\label{PI3k8}
PIP_3+PI3K_c&\stki{k_{PI3K_m}} PIP_3+PI3K_m \quad &\tn{on } \del\Omega\\
PI3K_c&\stk{\k{b,PI3K_m}}{k_{PI3K_c}} PI3K_m \quad &\tn{on } \del\Omega\\
Ras^*+PI3K_m&\stki{k_{PI3K_m^*}} Ras^*+PI3K_m^* \quad &\tn{on } \del\Omega\\
PI3K_m^*&\stki{k_{d,PI3K_m}} PI3K_m \quad &\tn{on } \del\Omega\\
PI3K_m^*+PIP_2&\stki{k_{PIP_3}} PI3K_m^*+PIP_3 \quad &\tn{on } \del\Omega\\
PTEN_m+PIP_3&\stki{k_{PIP_2}} PTEN_m+PIP_2 \quad &\tn{on } \del\Omega\\
PIP_2+PTEN_c&\stki{k_{PTEN_m}} PIP_2+PTEN_m \quad &\tn{on } \del\Omega\\
PTEN_m&\stki{k_{PTEN_c}} PTEN_c \quad &\tn{on } \del\Omega \label{PI3k15}
\end{align}
whose evolution can be described by a system of reaction-diffusion equations 
\begin{align*}
\delt{RasGEF^*}&=D_{RasGEF} \grad^2 RasGEF^* -\k{RasGEF} RasGEF^* \quad &\tn{in }\Omega\\
\delt{RasGAP^*}&=D_{RasGAP} \grad^2 RasGAP^* -\k{RasGAP} RasGAP^* \quad &\tn{in }\Omega\\
\delt{PI3K_c}&=D_{PI3K} \grad^2 PI3K_c \quad &\tn{in }\Omega\\
\delt{PTEN_c}&=D_{PTEN} \grad^2 PTEN_c \quad &\tn{in }\Omega\\
\delt{Ras^*}&=\lf(k_{Ras^*} RasGEF^*+\k{s,Ras^*}\rt) \cdot Ras-\lf(\k{Ras} RasGAP^* +\k{s,Ras}\rt)\cdot Ras^* \quad &\tn{on }\del\Omega\\
\delt{PI3K_m}&=\delta\k{b,PI3K_m} PI3K_c + k_{PI3K_m} PIP_3 \cdot PI3K_c + k_{d,PI3K_m} PI3K_m^* \\
&\qquad-k_{PI3K_m^*} Ras^* \cdot PI3K_m - k_{PI3K_c} PI3K_m \quad &\tn{on }\del\Omega\\
\delt{PI3K_m^*}&=k_{PI3K_m^*} Ras^* \cdot PI3K_m - k_{d,PI3K_m} PI3K_m^* \quad &\tn{on }\del\Omega\\
\delt{PTEN_m}&=\k{PTEN_m} PIP_2 \cdot PTEN_c - k_{PTEN_c} PTEN_m \quad &\tn{on }\del\Omega\\
\delt{PIP_3}&=k_{PIP_3} PI3K_m^* \cdot PIP_2 - \k{PIP_2} PTEN_m \cdot PIP_3 \quad &\tn{on }\del\Omega
\end{align*}
with the following boundary conditions for the cytosolic species
\begin{align*}
D_{RasGEF}\Dn{RasGEF^*}&=k_{RasGEF^*}\Gbg\cdot RasGEF \\
D_{RasGAP}\Dn{RasGAP^*}&=k_{RasGAP^*}\Gbg\cdot RasGAP \\
D_{PI3K}\Dn{PI3K_c}&=k_{PI3K_c} PI3K_m - k_{PI3K_m} PIP_3 \cdot PI3K_c - \delta\k{b,PI3K_m} PI3K_c \\
D_{PTEN}\Dn{PTEN_c}&=k_{PTEN_c} PTEN_m - \k{PTEN_m} PIP_2 \cdot PTEN_c 
\end{align*}
on $\del\Omega$, where $\del/\del n$ denotes the outward normal derivative, and conservation laws
\begin{align*}
RasGEF+RasGEF^*&=RasGEF_0\quad &\tn{in }\Omega\\
RasGAP+RasGAP^*&=RasGAP_0\quad &\tn{in }\Omega\\
Ras+Ras^*&=Ras_0\quad &\tn{on }\del\Omega\\
PIP_2+PIP_3&=P_0\quad &\tn{on }\del\Omega
\end{align*}
The justification for the form of the boundary conditions for RasGEF* and RasGAP* is given in
the following section. 

\subsection*{Simplification of membrane activation of cytosolic species}

We begin with a full description for RasGEF activity, which involves the 
spontaneous deactivation in the cytosol. 
\begin{align*}
\delt{RasGEF^*}&=D_{RasGEF} \grad^2 RasGEF^* -\k{RasGEF} RasGEF^* \\
\delt{RasGEF}&=D_{RasGEF} \grad^2 RasGEF +\k{RasGEF} RasGEF^* 
\end{align*}
in $\Omega$ with boundary conditions
\begin{align*}
D_{RasGEF}\Dn{RasGEF}&=-k_{b}\Gbg\cdot RasGEF + k_{ub}[\Gbg RasGEF] \\
D_{RasGEF}\Dn{RasGEF^*}&=k_{act}[\Gbg RasGEF] 
\end{align*}
on $\del\Omega$. These equations  account for binding to and activation by $\Gbg$ and the
dynamics of the membrane complex, given by 
\begin{align*}
\dx{[\Gbg RasGEF] }{t}&=k_{b}\Gbg\cdot RasGEF - (k_{ub}+\k{act})[\Gbg RasGEF] 
\end{align*}
on $\del\Omega$. If we assume that the complex is in a quasi steady state, i.e.
$\dx{[\Gbg RasGEF] }{t}=0$, then we have 
\begin{align*}
 [\Gbg RasGEF] =\frac{k_{b}}{k_{ub}+\k{act}}\Gbg\cdot RasGEF 
\end{align*}
and 
\begin{align*}
D_{RasGEF}\Dn{RasGEF}&=-D_{RasGEF}\Dn{RasGEF^*}=-\frac{k_{b}k_{act}}{k_{ub}+\k{
act}}\Gbg\cdot RasGEF 
\end{align*}
By defining $\k{RasGEF^*}=\dfrac{k_{b}k_{act}}{k_{ub}+\k{act}}$,  we obtain  the
simplified form used in the  model. A similar analysis applies to GAP*. 

 \subsection*{Parameters and details of the simulations}

Parameters used in the simulations are listed in Table \ref{PIPparamTable}. 
The surface densities of the membrane species in the model are taken from the
literature. Typical values of concentrations and diffusion constants, which are
0.1 $\mu M$ and 10 $\mu m^2/s$, respectively, are used for the cytosolic
species. The reaction-rate constants are chosen to match
experimentally-observed dynamics. In particular, the dynamics of the cytosolic
Ras-binding domain (RBD) reported in \cite{Takeda_2012} is used to match the
responses at different levels of uniform stimulation, while the responses to
static cAMP gradients are matched with the dynamics of PH$_\tn{Crac}$-GFP, a
PIP$_3$ reporter, at the front and the back of a live cell \cite{Xu_2005}. The
RBD dynamics is described by 
\begin{align*}
\delt{RBD_c}&=D_{RBD} \grad^2 RBD_c \quad &\tn{in }\Omega\\
D_{RBD}\Dn{RBD_c}&=k_{RBD_c} RBD_m - \k{RBD_m} Ras^* \cdot RBD_c \quad &\tn{on }\del\Omega\\
\delt{RBD_m}&=\k{RBD_m} Ras^* \cdot RBD_c - k_{RBD_c} RBD_m \quad &\tn{on }\del\Omega
\end{align*}
with the following parameters: $D_{RBD}=10~ \mu m^2/s,~RBD_0= 0.1~ \mu M,~k_{RBD_c}=7.5~ s^{-1}$, and $\k{RBD_m}=1200~\mu M^{-1} s^{-1}$.

The system is  solved numerically on two-dimensional domains by a finite element
method with backward differentiation formula for time stepping,  which is
implemented in the  COMSOL Multiphysics package. For each simulation, the system is
first simulated with uniform basal cAMP concentration until it reaches a steady
state. Then a stimulus is introduced by changing the external cAMP
profile. The cAMP level is represented by the surface density of free
G$_\beta\gamma$, which is the forcing function of the system. We assume that the
free G$_\beta\gamma$ density is proportional to the cAMP level and that half of
the heterotrimeric G protein on the membrane (with the total density of 2000
$\#/\mu m^2$ \cite{Meier-Schellersheim_2006}) is activated at 1 $\mu M$ cAMP.

\newpage
\subsection*{Adaptation of Ras activity with positive feedback from F-actin}

In this section we show that Ras activity still adapts under positive feedback from F-actin
under mild assumptions, given that the activation via cAMP and F-actin is
independent. Note that here we only give an analysis for the local dynamics
and assume that spontaneous activation and inactivation of Ras are negligible. 

\begin{figure}[h]
\centering
  \includegraphics[width=2.5 in]{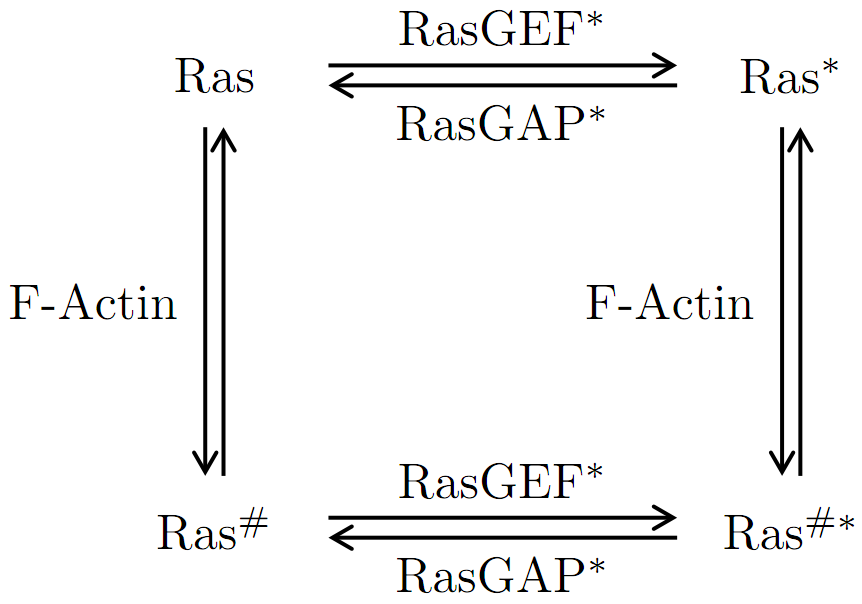}\\
\caption{{\bf An adapting Ras activation model with positive feedback from F-actin.} }\label{RasFB}
\end{figure}

Suppose
that Ras can be activated at two sites via cAMP and F-actin, respectively and
that these activations are independent, \ie the activation state at one site
does not affect the  activation/deactivation rates at the other site. The
activation diagram of Ras is shown in Figure \ref{RasFB} where $A$ is F-actin
concentration. 

The dynamics of Ras is given by 
\begin{align*}
\D\ea&=-\k\eb A\cdot\ea-\k\ec \gf\cdot\ea+\k{Ras,A}\eb+\k\ea \gp\cdot\ec\\
\D\eb&=\k\eb A\cdot\ea+\k\ea \gp\cdot\ed-\k{Ras,A}\eb-\k\ec \gf\cdot\eb\\
\D\ec&=\k \gf\gf\cdot\ea+\k{Ras,A}\ed-\k\ea\gp\cdot\ec-\k\eb A\cdot\ec\\
\D\ed&=-\k{Ras,A}\ed-\k\ea\gp\cdot\ed+\k\eb A\cdot\ec+\k\ec\gf\cdot\eb
\end{align*}
where we have conservation $\etot=\ea+\eb+\ec+\ed$. At steady state we have
\small\begin{align*}
\lf[\begin{array}{ccc}
\k{Ras,A}+\k\ec\gf&0&-\k\ea\gp\\
0&\k\ea\gp+\k\eb A&-\k{Ras,A}\\
-\k\ec\gf&-\k\eb A &\k{Ras,A}+\k\ea\gp
\end{array}\rt]
\lf[\begin{array}{c} \eb\\ \ec\\ \ed \end{array}\rt]
=\lf[\begin{array}{c} \k\eb A \cdot \ea\\ \k\ec\gf\cdot\ea\\ 0 \end{array}\rt]
\end{align*}\normalsize
So
\begin{align*}
\lf[\begin{array}{c} \eb\\ \ec\\ \ed \end{array}\rt]
=\ea\lf[\begin{array}{c} (\k\eb/\k{Ras,A})A\\ (\k\ec/\k\ea)\gf/\gp\\ (\k\eb\k\ec/\k{Ras,A}\k\ea)A\cdot\gf/\gp \end{array}\rt]
\end{align*}
and 
\begin{align*}
\ea=\frac{\etot}{\lf(1+(\k\eb/\k{Ras,A})A\rt)\lf( 1+(\k\ec/\k\ea)\gf/\gp\rt)}
\end{align*}
Assume that $\gf/\gp=c$ is a constant at steady state independently of cAMP concentration. Then 
\begin{align*}
\lf[\begin{array}{c} \ea\\\eb\\ \ec\\ \ed \end{array}\rt]
=\frac{\etot}{ \lf(1+(\k\eb/\k{Ras,A})A\rt)\lf( 1+(\k\ec/\k\ea)c\rt)}\lf[\begin{array}{c} 1\\(\k\eb/\k{Ras,A})A\\ (\k\ec/\k\ea)c\\ (\k \eb\k\ec/\k{Ras,A}\k\ea)cA \end{array}\rt]
\end{align*}
Therefore, a downstream activity of Ras $f(\ea,\eb,\ec\ed)$ is dependent on the positive feedback from F-actin but not on cAMP concentration.
	
\newpage
\subsection*{The effect of cell shape on activity of membrane-activated
proteins}

\subsubsection*{An infinite strip}

Consider activity of a protein which is activated at the boundary of a infinite
strip of $[0,L]\times\c{R}$ and spontaneously deactivated within the domain 
\begin{align*}
\delt{E^*}&=D\grad^2{E^*}-\k{1}E^*\tn{,\quad in $(0,L)\times\c R$}\\
D\Dn{E^*}&=\k{2}S(E_0-E^*)\tn{,\quad on $\{0, L\}\times\c R$}
\end{align*}
We want to solve for the steady-state enzyme activity at the membrane. First, we
can normalize the length $\theta=x/L$ so that the normalized domain is
$[0,1]\times\c R$ and $D$ and $\k 2$ are replaced by $D^+=D/L^2$ and $\k 2^+=\k
2 /L$. Since the problem is symmetric along the $y$-axis, $\del{E^*}/\del{y}=0$
and we can omit dependence on $y$. At the steady-state we have
\begin{align*}
D^+\Dxx{E^*}{\theta}&=\k{1}E^*\tn{,\quad in $\theta\in(0,L)$}
\end{align*}
whose solution is
\begin{align*}
E^*=c_1\cosh\lf(\theta/L^+_d\rt)+c_2\sinh\lf(\theta/L^+_d\rt)
\end{align*}
with
\begin{align*}
\Dx{E^*}{x}=\frac{1}{L^+_d}\lf(c_1\sinh\lf(\theta/L^+_d\rt)+c_2\cosh\lf(\theta/L^+_d\rt)\rt)
\end{align*}
where $L^+_d=\sqrt{\frac{ D^+}{\k1}}$ is normalized characteristic degradation
length. We assume symmetry across the midline,  and then the boundary conditions are  
\begin{align*}
-\frac{D^+}{L^+_d} c_2&=k_2^+S(E_0-c_1)\\
\frac{D^+}{L^+_d}\lf(c_1\sinh\lf(1/L^+_d\rt)+c_2\cosh\lf(1/L^+_d\rt)\rt)&=k_2^+S(E_0-c_1)
\end{align*}
at $\theta=0,1$ respectively. Hence
\begin{align*}
c_1&=E_0\lf/\lf(1+\frac{D^+}{k_2^+L^+_d S}\frac{ \sinh(1/L^+_d)}{1+\cosh(1/L^+_d)}\rt)\rt.\\
c_2&=-E_0\lf/\lf(\frac{1+\cosh(1/L^+_d)}{\sinh(1/L^+_d)}+\frac{D^+}{k_2^+L^+_d S}\rt)\rt.
\end{align*}
Define $L_a^+=\frac{D^+}{k_2^+ S}$ as normalized characteristic activation length. Then
\begin{align*}
E^*(\theta)=\lf(\frac{E_0}{1+\frac{L^+_a}{L^+_d}\frac{\sinh(1/L^+_d)}{1+\cosh(1/L^+_d)}}\rt)\lf(\cosh\lf(\theta/L^+_d\rt)-\frac{\sinh(1/L^+_d)}{1+\cosh(1/L^+_d)}\sinh (\theta/L^+_d)\rt)
\end{align*}
The activity at either boundary ($\theta=0,1$) is  
\begin{align*}
E^*=E_0\lf/\lf(1+\frac{L^+_a}{L^+_d}\frac{ \sinh(1/L^+_d)}{1+\cosh(1/L^+_d)}\rt)\rt.
\end{align*}

\subsubsection*{3D shell}
Next, consider activity of the membrane-activated protein in a 3D shell $\Omega=\lf\{\lf.(x,y,z)\in\c{R}^3\rt|l\leq x^2+y^2+z^2 \leq L\rt\}$. The activity is described by
\begin{align*}
\delt{E^*}&=D\grad^2{E^*}-\k{1}E^*\tn{,\quad in $\Omega$}\\
D\Dn{E^*}&=\k{2}S(E_0-E^*)\tn{,\quad on $\del\Omega$}
\end{align*}
In spherical coordinates, we have at steady state, by radial symmetry,
$$\frac{1}{r^2}\Dx{}{r}\lf(r^2\Dx{E^*}{r}\rt)=\k 1 E^*\tn{,\quad in $r\in(l,L)$}$$
Following \cite{Crank_1956}, let 
$$f=rE^*$$
so that
$$f_{rr}=\frac{\k1}{D}f=f/L_d^2,$$
where $L_d=\sqrt{D/k_1}$ is the characteristic degradation length, which leads to a general solution
\begin{align*}
E^*&=\frac{1}{r}\lf(c_1\cosh( r/L_d)+c_2\sinh(r/L_d)\rt)
\end{align*}
with 
\begin{align*}
\Dx{E^*}{r}&=\frac{1}{r^2}\lf(c_1\lf[\frac{r}{L_d} \sinh(r/L_d)-\cosh( r/L_d)\rt]+c_2\lf[\frac{ r}{L_d} \cosh(r/L_d)-\sinh(r/L_d)\rt]\rt)
\end{align*}
Recall boundary conditions 
\begin{align*}
-\Dx{E^*}{r}(l)&=\frac{1}{L_a}(E_0-E^*(l))\\
\Dx{E^*}{r}(L)&=\frac{1}{L_a}(E_0-E^*(L))
\end{align*}
where $L_a=D/\k{2}S$ is the characteristic activation length. Substitution gives 
\begin{align*}
c_1&=\frac{E_0}{L_a\Pi}\lf[ \frac{Ll}{L_d}(l\cosh( L/L_d)+L\cosh( l/L_d))+ \frac{Ll}{L_a}(l\sinh( L/L_d)-L\sinh( l/L_d))-l^2\sinh( L/L_d)-L^2\sinh( l/L_d)\rt]   \\
c_2&=\frac{ E_0}{L_a\Pi}\lf[-\frac{Ll}{L_d}(l\sinh( L/L_d)+L\sinh( l/L_d))-\frac{Ll}{L_a}(l\cosh( L/L_d)-L\cosh(l/L_d))+l^2\cosh(L/L_d)+L^2\cosh( l/L_d)\rt]
\end{align*}
where
\begin{align*}
\Pi&=\frac{ Ll}{L_d}\lf[\frac{1}{L_a}\cosh((L-l)/L_d)+\frac{1}{L_d}\sinh((L-l)/L_d)\rt]
+\frac{Ll}{L_a}\lf[\frac{1}{L_d}\cosh((L-l)/L_d)+\frac{1}{L_a}\sinh((L-l)/L_d)\rt]\\
&\qquad+(L-l)\lf[\frac{1}{L_d}\cosh((L-l)/L_d)+\frac{1}{L_a}\sinh((L-l)/L_d)\rt]-\sinh((L-l)/L_d)
\end{align*}
and 
\begin{align*}
E^*(r)&=\frac{E_0}{L_a\Pi r}\lf\{\frac{ Ll}{L_d}\lf[l\cosh((L-r)/L_d)+L\cosh((r-l)/L_d)\rt]
+\frac{Ll}{L_a}\lf[l\sinh((L-r)/L_d)+L\sinh((r-l)/L_d)\rt]\rt.\\
&\quad\lf.\phantom{\frac{Ll}{L_a}}-l^2\sinh((L-r)/L_d)+L^2\sinh((r-l)/L_d)
\rt\}
\end{align*}
The expression can be simplified as 
\begin{align*}\small
E^*(r)=\lf\{\begin{aligned}
&\frac{ E_0}{L_ar}\frac{KLl[l\cosh((L-r)/L_d+\phi)+L\cosh((r-l)/L_d+\phi)]-l^2\sinh((L-r)/L_d)+L^2\sinh((r-l)/L_d)}{K^2Ll\sinh((L-l)/L_d+2\phi)+K(L-l)\cosh((L-l)/L_d+\phi)-\sinh((L-l)/L_d)} & \text{for }L_d<L_a \\
&\frac{ E_0}{L_ar}\frac{ (Ll/L_d)[le^{(L-r)/L_d}+Le^{(r-l)/L_d}]-l^2\sinh((L-r)/L_d)+L^2\sinh((r-l)/L_d)}
{2 (Ll/L_d^2)e^{(L-l)/L_d}
+((L-l)/L_d)e^{(L-l)/L_d}-\sinh((L-l)/L_d)} & \text{for }L_d=L_a  \\
&\frac{ E_0}{L_ar}\frac{KLl[l\sinh((L-r)/L_d+\phi)+L\sinh((r-l)/L_d+\phi)]-l^2\sinh((L-r)/L_d)+L^2\sinh((r-l)/L_d)}{K^2Ll\sinh((L-l)/L_d+2\phi)+K(L-l)\sinh((L-l)/L_d+\phi)-\sinh((L-l)/L_d)} & \text{for }L_d>L_a 
\end{aligned}\rt.\normalsize
\end{align*}
where 
\begin{align*}
K=\sqrt{\lf|\frac{1}{L_d^2}-\frac{1}{L_a^2}\rt|}\qquad \text{and} \qquad \phi=
\begin{cases}
\tanh^{-1}(L_d/L_a) & \text{for }L_d<L_a \\
\tanh^{-1}(L_a/L_d) & \text{for }L_d>L_a 
\end{cases}
\end{align*}

\section*{Acknowledgements} We are grateful
to Benjamin Jordan for a careful critique of the manuscript.

\FloatBarrier
\section*{Supporting information} 

{\bf Figure S1. $\mathbf{PIP_3}$ localization dynamics under uniform stimulation.} A resting cell 
is subject to 0.1 $\mu M$ cAMP at 0 $s$.
\\\\
\noindent {\bf Figure S2. $\mathbf{PIP_3}$ localization dynamics under an upward gradient.} A resting 
cell is subject to a directional cue in the direction of the \textit{y}-axis with 50\% difference 
in cAMP levels across the cell. The mean cAMP level is 0.1 $\mu M$.


\begin{thebibliography}{10}
\providecommand{\url}[1]{
}
\providecommand{\urlprefix}{
}
\expandafter\ifx\csname urlstyle\endcsname\relax
  \providecommand{\doi}[1]{doi:\discretionary{}{}{}#1}\else
  \providecommand{\doi}{doi:\discretionary{}{}{}\begingroup
  \urlstyle{rm}\Url}\fi
\providecommand{\bibAnnoteFile}[1]{%
}
\providecommand{\bibAnnote}[2]{%
}
\providecommand{\eprint}[2][]{\url{#2}}

\bibitem{Othmer:2013:EAB}
Othmer HG, Xin X, Xue C (2013) Excitation and adaptation in bacteria--a model
  signal transduction system that controls taxis and spatial pattern formation.
\newblock International journal of molecular sciences 14: 9205--9248.
\bibAnnoteFile{Othmer:2013:EAB}

\bibitem{Othmer:1998:OCS}
Othmer H, Schaap P (1998) Oscillatory c{AMP} signaling in the development of
  {D}ictyostelium discoideum.
\newblock Comments on Theoretical Biology 5: 175--282.
\bibAnnoteFile{Othmer:1998:OCS}

\bibitem{Soll:1993:MBA}
Soll DR, Wessels D, Sylwester A (1993) The motile behavior of amoebae in the
  aggregation wave in {\em \uppercase{d}\lowercase{ictyostelium discoideum}}.
\newblock In: Othmer HG, Maini PK, Murray JD, editors, Experimental and
  Theoretical Advances in Biological Pattern Formation. London: Plenum, pp.
  325--338.
\bibAnnote{Soll:1993:MBA}{Dd, cell motion}

\bibitem{Soll:1995:UCU}
Soll DR (1995) The use of computers in understanding how animal cells crawl.
\newblock In: Jeon KW, Jarvik J, editors, International Review of Cytology,
  Acdemic Press, volume 163. pp. 43--104.
\bibAnnote{Soll:1995:UCU}{chemotaxis}

\bibitem{Haastert_2004}
Haastert PJMV, Devreotes PN (2004) {C}hemotaxis: signalling the way forward.
\newblock Nat Rev Mol Cell Biol 5: 626--634.
\bibAnnoteFile{Haastert_2004}

\bibitem{Li_2008}
Li L, Nørrelykke SF, Cox EC (2008) {P}ersistent cell motion in the absence of
  external signals: a search strategy for eukaryotic cells.
\newblock PLoS One 3: e2093.
\bibAnnoteFile{Li_2008}

\bibitem{Bosgraaf_2009}
Bosgraaf L, Haastert PJMV (2009) {T}he ordered extension of pseudopodia by
  amoeboid cells in the absence of external cues.
\newblock PLoS One 4: e5253.
\bibAnnoteFile{Bosgraaf_2009}

\bibitem{McRobbie:1983:CAA}
McRobbie S, Newell P (1983) Changes in actin associated with the cytoskeleton
  following chemotactic stimulation of dictyostelium discoideum.
\newblock Biochemical and biophysical research communications 115: 351--359.
\bibAnnoteFile{McRobbie:1983:CAA}

\bibitem{Condeelis:1990:MAC}
Condeelis J, et~al. (1990) Mechanisms of ameboid chemotaxis: An evaluation of
  the cortical expansion model.
\newblock Developmental Genetics 11: 333--340.
\bibAnnote{Condeelis:1990:MAC}{cell movement, chemotaxis}

\bibitem{Varnum:1985:DAA}
Varnum B, Edwards KB, Soll DR (1985) {{\em {Dictyostelium}\/}} amebae alter
  motility differently in response to increasing versus decreasing temporal
  gradients of {cAMP} 101: 1--5.
\bibAnnote{Varnum:1985:DAA}{Dictyostelium}

\bibitem{Wessels:1992:BDA}
Wessels D, Murray J, Soll DR (1992) Behavior of {{\em {Dictyostelium}\/}}
  amoebae is regulated primarily by the temporal dynamic of the natural {cAMP}
  wave.
\newblock Cell Motil Cytoskeleton 23: 145--156.
\bibAnnote{Wessels:1992:BDA}{pattern formation}

\bibitem{Zigmond_1977}
Zigmond SH (1977) Ability of polymorphonuclear leukocytes to orient in
  gradients of chemotactic factors.
\newblock J Cell Biol 75: 606--616.
\bibAnnoteFile{Zigmond_1977}

\bibitem{Kriebel_2003}
Kriebel PW, Barr VA, Parent CA (2003) {A}denylyl cyclase localization regulates
  streaming during chemotaxis.
\newblock Cell 112: 549--560.
\bibAnnoteFile{Kriebel_2003}

\bibitem{Swanson:1982:LSC}
Swanson J, Taylor DL (1982) Local and spatially coordinated movements in {{\em
  {D}}ictyostelium discoideum} amoedae during chemotaxis 28: 225--232.
\bibAnnote{Swanson:1982:LSC}{Dd}

\bibitem{Alcantara:1974:SPD}
Alcantara F, Monk M (1974) Signal propagation during aggregation in the slime
  mold {\em \uppercase{d}\lowercase{ictyostelium discoideum}}.
\newblock J Gen Microbiol 85: 321--334.
\bibAnnoteFile{Alcantara:1974:SPD}

\bibitem{Meier_2011}
Meier B, Zielinski A, Weber C, Arcizet D, Youssef S, et~al. (2011)
  {C}hemotactic cell trapping in controlled alternating gradient fields.
\newblock Proc Natl Acad Sci U S A 108: 11417--11422.
\bibAnnoteFile{Meier_2011}

\bibitem{Parent:1999:CSD}
Parent CA, Devreotes PN (1999) A cell's sense of direction.
\newblock Science 284: 765--770.
\bibAnnote{Parent:1999:CSD}{chemotaxis, Dd}

\bibitem{Gerisch:1982:CD}
Gerisch G (1982) Chemotaxis in {{\em {Dictyostelium}\/}} 44: 535--552.
\bibAnnote{Gerisch:1982:CD}{dictyostelium,chemotaxis,cell movement}

\bibitem{Andrew_2007}
Andrew N, Insall RH (2007) {C}hemotaxis in shallow gradients is mediated
  independently of {P}td{I}ns 3-kinase by biased choices between random
  protrusions.
\newblock Nat Cell Biol 9: 193--200.
\bibAnnoteFile{Andrew_2007}

\bibitem{Fisher:1989:QAC}
Fisher PR, Merkl R, Gerisch G (1989) Quantitative analysis of cell motility and
  chemotaxis in {{\it {D}ictyostelium discoideum}} by using an image processing
  system and a novel chemotaxis chamber providing stationary chemical
  gradients.
\newblock J Cell Biol 108: 973--984.
\bibAnnote{Fisher:1989:QAC}{cell motion, dictyostelium, chemotaxis}

\bibitem{Brenner:1984:CBA}
Brenner M, Thomas SD (1984) Caffeine blocks activation of cyclic {AMP}
  synthesis in {\em {dictyostelium} discoideum}.
\newblock Dev Biol 101: 136--146.
\bibAnnote{Brenner:1984:CBA}{dictyostelium}

\bibitem{Siegert:1989:DIP}
Siegert F, Weijer CJ (1989) Digital image processing of optical density wave
  propagation in {\em \uppercase{d}\lowercase{ictyostelium discoideum}} and
  analysis of the effects of caffeine and ammonia 93: 325--335.
\bibAnnote{Siegert:1989:DIP}{Dictyostelium discoideum; waves}

\bibitem{Parent_1998}
Parent CA, Blacklock BJ, Froehlich WM, Murphy DB, Devreotes PN (1998) {G}
  protein signaling events are activated at the leading edge of chemotactic
  cells.
\newblock Cell 95: 81--91.
\bibAnnoteFile{Parent_1998}

\bibitem{Takeda_2012}
Takeda K, Shao D, Adler M, Charest PG, Loomis WF, et~al. (2012) Incoherent
  feedforward control governs adaptation of activated {R}as in a eukaryotic
  chemotaxis pathway.
\newblock Sci Signal 5: ra2.
\bibAnnoteFile{Takeda_2012}

\bibitem{Tang_1995}
Tang Y, Othmer HG (1995) {E}xcitation, oscillations and wave propagation in a
  {G}-protein-based model of signal transduction in {D}ictyostelium discoideum.
\newblock Philos Trans R Soc Lond B Biol Sci 349: 179--195.
\bibAnnoteFile{Tang_1995}

\bibitem{Dallon_1998}
Dallon JC, Othmer HG (1998) {A} continuum analysis of the chemotactic signal
  seen by {D}ictyostelium discoideum.
\newblock J Theor Biol 194: 461--483.
\bibAnnoteFile{Dallon_1998}

\bibitem{Chen_2003}
Chen L, Janetopoulos C, Huang YE, Iijima M, Borleis J, et~al. (2003) {T}wo
  phases of actin polymerization display different dependencies on
  {P}{I}(3,4,5){P}3 accumulation and have unique roles during chemotaxis.
\newblock Mol Biol Cell 14: 5028--5037.
\bibAnnoteFile{Chen_2003}

\bibitem{Sasaki_2004}
Sasaki AT, Chun C, Takeda K, Firtel RA (2004) {L}ocalized {R}as signaling at
  the leading edge regulates {P}{I}3{K}, cell polarity, and directional cell
  movement.
\newblock J Cell Biol 167: 505--518.
\bibAnnoteFile{Sasaki_2004}

\bibitem{Xu_2005}
Xu X, Meier-Schellersheim M, Jiao X, Nelson LE, Jin T (2005) {Q}uantitative
  imaging of single live cells reveals spatiotemporal dynamics of multistep
  signaling events of chemoattractant gradient sensing in {D}ictyostelium.
\newblock Mol Biol Cell 16: 676--688.
\bibAnnoteFile{Xu_2005}

\bibitem{Kortholt_2011}
Kortholt A, Kataria R, Keizer-Gunnink I, Egmond WNV, Khanna A, et~al. (2011)
  {D}ictyostelium chemotaxis: essential {R}as activation and accessory
  signalling pathways for amplification.
\newblock EMBO Rep 12: 1273--1279.
\bibAnnoteFile{Kortholt_2011}

\bibitem{Zhang_2008}
Zhang S, Charest PG, Firtel RA (2008) {S}patiotemporal regulation of {R}as
  activity provides directional sensing.
\newblock Curr Biol 18: 1587--1593.
\bibAnnoteFile{Zhang_2008}

\bibitem{Funamoto_2001}
Funamoto S, Milan K, Meili R, Firtel RA (2001) {R}ole of phosphatidylinositol
  3' kinase and a downstream pleckstrin homology domain-containing protein in
  controlling chemotaxis in dictyostelium.
\newblock J Cell Biol 153: 795--810.
\bibAnnoteFile{Funamoto_2001}

\bibitem{Afonso_2011}
Afonso PV, Parent CA (2011) Pi3k and chemotaxis: a priming issue?
\newblock Sci Signal 4: pe22.
\bibAnnoteFile{Afonso_2011}

\bibitem{Sasaki_2007}
Sasaki AT, Janetopoulos C, Lee S, Charest PG, Takeda K, et~al. (2007) {G}
  protein-independent {R}as/{P}{I}3{K}/{F}-actin circuit regulates basic cell
  motility.
\newblock J Cell Biol 178: 185--191.
\bibAnnoteFile{Sasaki_2007}

\bibitem{Welf_2012}
Welf ES, Ahmed S, Johnson HE, Melvin AT, Haugh JM (2012) Migrating fibroblasts
  reorient directionality by a metastable, pi3k-dependent mechanism.
\newblock J Cell Biol 197: 105--114.
\bibAnnoteFile{Welf_2012}

\bibitem{Goldbeter_1977}
Goldbeter A, Segel LA (1977) Unified mechanism for relay and oscillation of
  cyclic amp in dictyostelium discoideum.
\newblock Proc Natl Acad Sci U S A 74: 1543--1547.
\bibAnnoteFile{Goldbeter_1977}

\bibitem{Meinhardt_1999}
Meinhardt H (1999) {O}rientation of chemotactic cells and growth cones: models
  and mechanisms.
\newblock J Cell Sci 112 ( Pt 17): 2867--2874.
\bibAnnoteFile{Meinhardt_1999}

\bibitem{Meier-Schellersheim_2006}
Meier-Schellersheim M, Xu X, Angermann B, Kunkel EJ, Jin T, et~al. (2006) {K}ey
  role of local regulation in chemosensing revealed by a new molecular
  interaction-based modeling method.
\newblock PLoS Comput Biol 2: e82.
\bibAnnoteFile{Meier-Schellersheim_2006}

\bibitem{Xiong_2010}
Xiong Y, Huang CH, Iglesias PA, Devreotes PN (2010) {C}ells navigate with a
  local-excitation, global-inhibition-biased excitable network.
\newblock Proc Natl Acad Sci U S A 107: 17079--17086.
\bibAnnoteFile{Xiong_2010}

\bibitem{Gierer:1972:TBP}
Gierer A, Meinhardt H (1972) A theory of biological pattern formation.
\newblock Biological Cybernetics 12: 30--39.
\bibAnnoteFile{Gierer:1972:TBP}

\bibitem{Shi_2013}
Shi C, Huang CH, Devreotes PN, Iglesias PA (2013) Interaction of motility,
  directional sensing, and polarity modules recreates the behaviors of
  chemotaxing cells.
\newblock PLOS Computational Biology 9: e1003122.
\bibAnnoteFile{Shi_2013}

\bibitem{Gamba_2005}
Gamba A, de~Candia A, Talia SD, Coniglio A, Bussolino F, et~al. (2005)
  {D}iffusion-limited phase separation in eukaryotic chemotaxis.
\newblock Proc Natl Acad Sci U S A 102: 16927--16932.
\bibAnnoteFile{Gamba_2005}

\bibitem{Mori_2008}
Mori Y, Jilkine A, Edelstein-Keshet L (2008) {W}ave-pinning and cell polarity
  from a bistable reaction-diffusion system.
\newblock Biophys J 94: 3684--3697.
\bibAnnoteFile{Mori_2008}

\bibitem{Hecht_2010}
Hecht I, Kessler DA, Levine H (2010) {T}ransient localized patterns in
  noise-driven reaction-diffusion systems.
\newblock Phys Rev Lett 104: 158301.
\bibAnnoteFile{Hecht_2010}

\bibitem{Jilkine_2011}
Jilkine A, Edelstein-Keshet L (2011) A comparison of mathematical models for
  polarization of single eukaryotic cells in response to guided cues.
\newblock PLoS Comput Biol 7: e1001121.
\bibAnnoteFile{Jilkine_2011}

\bibitem{Levchenko_2002}
Levchenko A, Iglesias PA (2002) {M}odels of eukaryotic gradient sensing:
  application to chemotaxis of amoebae and neutrophils.
\newblock Biophys J 82: 50--63.
\bibAnnoteFile{Levchenko_2002}

\bibitem{Kortholt_2008}
Kortholt A, van Haastert PJM (2008) {H}ighlighting the role of {R}as and {R}ap
  during {D}ictyostelium chemotaxis.
\newblock Cell Signal 20: 1415--1422.
\bibAnnoteFile{Kortholt_2008}

\bibitem{Kataria_2013}
Kataria R, Xu X, Fusetti F, Keizer-Gunnink I, Jin T, et~al. (2013)
  Dictyostelium ric8 is a nonreceptor guanine exchange factor for
  heterotrimeric g proteins and is important for development and chemotaxis.
\newblock Proc Natl Acad Sci U S A 110: 6424--6429.
\bibAnnoteFile{Kataria_2013}

\bibitem{Wilkins:2005:DGE}
Wilkins A, Szafranski K, Fraser DJ, Bakthavatsalam D, Muller R, et~al. (2005)
  The dictyostelium genome encodes numerous ras{GEF}s with multiple biological
  roles.
\newblock Genome Biol 6: R68.
\bibAnnote{Wilkins:2005:DGE}{dictyostelium}

\bibitem{Kae_2007}
Kae H, Kortholt A, Rehmann H, Insall RH, Haastert PJMV, et~al. (2007) {C}yclic
  {A}{M}{P} signalling in {D}ictyostelium: {G}-proteins activate separate {R}as
  pathways using specific {R}as{G}{E}{F}s.
\newblock EMBO Rep 8: 477--482.
\bibAnnoteFile{Kae_2007}

\bibitem{Funamoto_2002}
Funamoto S, Meili R, Lee S, Parry L, Firtel RA (2002) {S}patial and temporal
  regulation of 3-phosphoinositides by {P}{I} 3-kinase and {P}{T}{E}{N}
  mediates chemotaxis.
\newblock Cell 109: 611--623.
\bibAnnoteFile{Funamoto_2002}

\bibitem{Bolourani_2008}
Bolourani P, Spiegelman GB, Weeks G (2008) {R}ap1 activation in response to
  c{A}{M}{P} occurs downstream of ras activation during {D}ictyostelium
  aggregation.
\newblock J Biol Chem 283: 10232--10240.
\bibAnnoteFile{Bolourani_2008}

\bibitem{Charest_2010}
Charest PG, Shen Z, Lakoduk A, Sasaki AT, Briggs SP, et~al. (2010) {A} {R}as
  signaling complex controls the {R}as{C}-{T}{O}{R}{C}2 pathway and directed
  cell migration.
\newblock Dev Cell 18: 737--749.
\bibAnnoteFile{Charest_2010}

\bibitem{Park_2004}
Park KC, Rivero F, Meili R, Lee S, Apone F, et~al. (2004) {R}ac regulation of
  chemotaxis and morphogenesis in {D}ictyostelium.
\newblock EMBO J 23: 4177--4189.
\bibAnnoteFile{Park_2004}

\bibitem{Han_2006}
Han JW, Leeper L, Rivero F, Chung CY (2006) {R}ole of {R}ac{C} for the
  regulation of {W}{A}{S}{P} and phosphatidylinositol 3-kinase during
  chemotaxis of {D}ictyostelium.
\newblock J Biol Chem 281: 35224--35234.
\bibAnnoteFile{Han_2006}

\bibitem{Yan_2012}
Yan J, Mihaylov V, Xu X, Brzostowski JA, Li H, et~al. (2012) {A}
  {G}$\beta\gamma$ {E}ffector, {E}lmo{E}, {T}ransduces {G}{P}{C}{R} {S}ignaling
  to the {A}ctin {N}etwork during {C}hemotaxis.
\newblock Dev Cell 22: 92--103.
\bibAnnoteFile{Yan_2012}

\bibitem{Loovers_2006}
Loovers HM, Postma M, Keizer-Gunnink I, Huang YE, Devreotes PN, et~al. (2006)
  {D}istinct roles of {P}{I}(3,4,5){P}3 during chemoattractant signaling in
  {D}ictyostelium: a quantitative in vivo analysis by inhibition of
  {P}{I}3-kinase.
\newblock Mol Biol Cell 17: 1503--1513.
\bibAnnoteFile{Loovers_2006}

\bibitem{Kortholt_2006}
Kortholt A, Rehmann H, Kae H, Bosgraaf L, Keizer-Gunnink I, et~al. (2006)
  {C}haracterization of the {G}bp{D}-activated {R}ap1 pathway regulating
  adhesion and cell polarity in {D}ictyostelium discoideum.
\newblock J Biol Chem 281: 23367--23376.
\bibAnnoteFile{Kortholt_2006}

\bibitem{Jeon_2007}
Jeon TJ, Lee DJ, Lee S, Weeks G, Firtel RA (2007) {R}egulation of {R}ap1
  activity by {R}ap{G}{A}{P}1 controls cell adhesion at the front of
  chemotaxing cells.
\newblock J Cell Biol 179: 833--843.
\bibAnnoteFile{Jeon_2007}

\bibitem{Kortholt_2010}
Kortholt A, Bolourani P, Rehmann H, Keizer-Gunnink I, Weeks G, et~al. (2010)
  {A} {R}ap/phosphatidylinositol 3-kinase pathway controls pseudopod formation.
\newblock Mol Biol Cell 21: 936--945.
\bibAnnoteFile{Kortholt_2010}

\bibitem{Janetopoulos_2001}
Janetopoulos C, Jin T, Devreotes P (2001) {R}eceptor-mediated activation of
  heterotrimeric {G}-proteins in living cells.
\newblock Science 291: 2408--2411.
\bibAnnoteFile{Janetopoulos_2001}

\bibitem{Xu_2007}
Xu X, Meier-Schellersheim M, Yan J, Jin T (2007) {L}ocally controlled
  inhibitory mechanisms are involved in eukaryotic {G}{P}{C}{R}-mediated
  chemosensing.
\newblock J Cell Biol 178: 141--153.
\bibAnnoteFile{Xu_2007}

\bibitem{Khamviwath:2013:MAS}
Khamviwath V, Othmer HG (2013) Modular analtsis of signal transduction
  networks.
\newblock In preparation.
\bibAnnoteFile{Khamviwath:2013:MAS}

\bibitem{Weiner_2002}
Weiner OD, Neilsen PO, Prestwich GD, Kirschner MW, Cantley LC, et~al. (2002) A
  ptdinsp(3)- and rho gtpase-mediated positive feedback loop regulates
  neutrophil polarity.
\newblock Nat Cell Biol 4: 509--513.
\bibAnnoteFile{Weiner_2002}

\bibitem{Postma_2004}
Postma M, Roelofs J, Goedhart J, Loovers HM, Visser AJWG, et~al. (2004)
  {S}ensitization of {D}ictyostelium chemotaxis by
  phosphoinositide-3-kinase-mediated self-organizing signalling patches.
\newblock J Cell Sci 117: 2925--2935.
\bibAnnoteFile{Postma_2004}

\bibitem{Ferguson_2007}
Ferguson GJ, Milne L, Kulkarni S, Sasaki T, Walker S, et~al. (2007) Pi(3)kgamma
  has an important context-dependent role in neutrophil chemokinesis.
\newblock Nat Cell Biol 9: 86--91.
\bibAnnoteFile{Ferguson_2007}

\bibitem{Khamviwath_2013}
Khamviwath V, Hu J, Othmer HG (2013) A continuum model of actin waves in
  dictyostelium discoideum.
\newblock PLoS One 8: e64272.
\bibAnnoteFile{Khamviwath_2013}

\bibitem{Hoeller_2007}
Hoeller O, Kay RR (2007) {C}hemotaxis in the absence of {P}{I}{P}3 gradients.
\newblock Curr Biol 17: 813--817.
\bibAnnoteFile{Hoeller_2007}

\bibitem{Schneider_2005}
Schneider IC, Parrish EM, Haugh JM (2005) Spatial analysis of 3'
  phosphoinositide signaling in living fibroblasts, iii: influence of cell
  morphology and morphological polarity.
\newblock Biophys J 89: 1420--1430.
\bibAnnoteFile{Schneider_2005}

\bibitem{Krishnan_2007}
Krishnan J, Iglesias PA (2007) {R}eceptor-mediated and intrinsic polarization
  and their interaction in chemotaxing cells.
\newblock Biophys J 92: 816--830.
\bibAnnoteFile{Krishnan_2007}

\bibitem{Hecht_2011}
Hecht I, Skoge ML, Charest PG, Ben-Jacob E, Firtel RA, et~al. (2011)
  {A}ctivated membrane patches guide chemotactic cell motility.
\newblock PLoS Comput Biol 7: e1002044.
\bibAnnoteFile{Hecht_2011}

\bibitem{Wang_2012}
Wang CJ, Bergmann A, Lin B, Kim K, Levchenko A (2012) Diverse sensitivity
  thresholds in dynamic signaling responses by social amoebae.
\newblock Sci Signal 5: ra17.
\bibAnnoteFile{Wang_2012}

\bibitem{Dinauer_1980}
Dinauer MC, Steck TL, Devreotes PN (1980) {C}yclic 3',5'-{A}{M}{P} relay in
  {D}ictyostelium discoideum {I}{V}. {R}ecovery of the c{A}{M}{P} signaling
  response after adaptation to c{A}{M}{P}.
\newblock J Cell Biol 86: 545--553.
\bibAnnoteFile{Dinauer_1980}

\bibitem{Machacek_2006}
Machacek M, Danuser G (2006) Morphodynamic profiling of protrusion phenotypes.
\newblock Biophys J 90: 1439--1452.
\bibAnnoteFile{Machacek_2006}

\bibitem{Houk_2012}
Houk AR, Jilkine A, Mejean CO, Boltyanskiy R, Dufresne ER, et~al. (2012)
  Membrane tension maintains cell polarity by confining signals to the leading
  edge during neutrophil migration.
\newblock Cell 148: 175--188.
\bibAnnoteFile{Houk_2012}

\bibitem{Yoo_2010}
Yoo SK, Deng Q, Cavnar PJ, Wu YI, Hahn KM, et~al. (2010) Differential
  regulation of protrusion and polarity by pi3k during neutrophil motility in
  live zebrafish.
\newblock Dev Cell 18: 226--236.
\bibAnnoteFile{Yoo_2010}

\bibitem{Gerisch_2010}
Gerisch G (2010) {S}elf-organizing actin waves that simulate phagocytic cup
  structures.
\newblock PMC Biophys 3: 7.
\bibAnnoteFile{Gerisch_2010}

\bibitem{Arai_2010}
Arai Y, Shibata T, Matsuoka S, Sato MJ, Yanagida T, et~al. (2010)
  {S}elf-organization of the phosphatidylinositol lipids signaling system for
  random cell migration.
\newblock Proc Natl Acad Sci U S A 107: 12399--12404.
\bibAnnoteFile{Arai_2010}

\bibitem{Taniguchi_2013}
Taniguchi D, Ishihara S, Oonuki T, Honda-Kitahara M, Kaneko K, et~al. (2013)
  Phase geometries of two-dimensional excitable waves govern self-organized
  morphodynamics of amoeboid cells.
\newblock Proc Natl Acad Sci U S A 110: 5016--5021.
\bibAnnoteFile{Taniguchi_2013}

\bibitem{Bosgraaf_2008}
Bosgraaf L, Keizer-Gunnink I, Haastert PJMV (2008) {P}{I}3-kinase signaling
  contributes to orientation in shallow gradients and enhances speed in steep
  chemoattractant gradients.
\newblock J Cell Sci 121: 3589--3597.
\bibAnnoteFile{Bosgraaf_2008}

\bibitem{Comer_2005}
Comer FI, Lippincott CK, Masbad JJ, Parent CA (2005) {T}he {P}{I}3{K}-mediated
  activation of {C}{R}{A}{C} independently regulates adenylyl cyclase
  activation and chemotaxis.
\newblock Curr Biol 15: 134--139.
\bibAnnoteFile{Comer_2005}

\bibitem{Yang_2007}
Yang C, Czech L, Gerboth S, ichiro Kojima S, Scita G, et~al. (2007) {N}ovel
  roles of formin m{D}ia2 in lamellipodia and filopodia formation in motile
  cells.
\newblock PLoS Biol 5: e317.
\bibAnnoteFile{Yang_2007}

\bibitem{Bosgraaf_2009a}
Bosgraaf L, Haastert PJMV (2009) {N}avigation of chemotactic cells by parallel
  signaling to pseudopod persistence and orientation.
\newblock PLoS One 4: e6842.
\bibAnnoteFile{Bosgraaf_2009a}

\bibitem{VanHaastert:1984:DRH}
{{Van Haastert}} PJM, de~Wit RJ (1984) Demonstration of receptor heterogeneity
  and affinity modulation by nonequilibrium binding experiments.
\newblock J of Biological Chemistry 259: 13321--13328.
\bibAnnote{VanHaastert:1984:DRH}{Dd}

\bibitem{Berg:1977:PC}
Berg HC, Purcell EM (1977) Physics of chemoreception.
\newblock Biophys J 20: 193--219.
\bibAnnote{Berg:1977:PC}{pattern formation}

\bibitem{VanHaastert:1987:RCS}
{Van Haastert} PJ (1987) Down-regulation of cell surface cyclic {AMP} receptors
  and desensitization of cyclic {AMP-stimulated} adenylate cyclase by cyclic
  {AMP} in {{\em {Dictyostelium} discoideum\/}}. kinetics and concentration
  dependence 262: 7700--7704.
\bibAnnoteFile{VanHaastert:1987:RCS}

\bibitem{Eigen:1963:AE}
Eigen M, Hammes G (1963) Advances in Enztmology, Interscience, volume~25,
  chapter Elementary steps in enzyme reactions.
\newblock pp. 1--38.
\bibAnnote{Eigen:1963:AE}{diffusion controlled reactions}

\bibitem{Ma_2004}
Ma L, Janetopoulos C, Yang L, Devreotes PN, Iglesias PA (2004) {T}wo
  complementary, local excitation, global inhibition mechanisms acting in
  parallel can explain the chemoattractant-induced regulation of
  {P}{I}(3,4,5){P}3 response in dictyostelium cells.
\newblock Biophys J 87: 3764--3774.
\bibAnnoteFile{Ma_2004}

\bibitem{Postma_2001}
Postma M, {Van Haastert} PJ (2001) {A} diffusion-translocation model for
  gradient sensing by chemotactic cells.
\newblock Biophys J 81: 1314--1323.
\bibAnnoteFile{Postma_2001}

\bibitem{Crank_1956}
Crank J (1956) The Mathematics of Diffusion.
\newblock Clarendon Press.
\bibAnnoteFile{Crank_1956}

\end{thebibliography}
\end{document}